\documentclass[a4paper,11pt]{article}
\pdfoutput=1
\usepackage{jinstpub-mod}

\usepackage[sort&compress]{natbib}
\usepackage{booktabs}

\title{Development of a Bonner Sphere Neutron Spectrometer from a Commercial Neutron Dosimeter}

\author[a]{M.~C.~Chu}
\author[b]{K.~Y.~Fung}
\author[b]{T.~Kwok}
\author[b]{J.~K.~C.~Leung}
\author[a,b]{Y.~C.~Lin}
\author[b,c]{H.~Liu}
\author[d]{K.~B.~Luk}
\author[b,1]{H.~Y.~Ngai,\note{Corresponding author.}}
\author[b]{C.~S.~J.~Pun}
\author[b,d]{H.~L.~H.~Wong}

\affiliation[a]{Department of Physics, The Chinese University of Hong Kong,\\Shatin, N.T., Hong Kong}
\affiliation[b]{Department of Physics, The University of Hong Kong,\\Pokfulam, Hong Kong}
\affiliation[c]{Department of Physics, University of Virginia,\\Charlottesville, VA 22904, USA}
\affiliation[d]{Department of Physics, University of California at Berkeley,\\Berkeley, CA 94720, USA}

\emailAdd{jngai@graduate.hku.hk}

\abstract{Bonner Spheres have been used widely for the measurement of neutron spectra with neutron energies ranged from thermal up to at least 20 MeV. A Bonner Sphere neutron spectrometer (BSS) was developed by extending a Berthold LB 6411 neutron-dose-rate meter. The BSS consists of a $^{3}$He thermal-neutron detector with integrated electronics, a set of eight polyethylene spherical shells and two optional lead shells of various sizes. The response matrix of the BSS was calculated with GEANT4 Monte Carlo simulation. The BSS had a calibration uncertainty of $\pm 8.6\%$ and a detector background rate of $(1.57 \pm 0.04) \times 10^{-3}$ s$^{-1}$. A spectral unfolding code NSUGA was developed. The NSUGA code utilizes genetic algorithms and has been shown to perform well in the absence of \textit{a priori} information.\note{This is an author-created, un-copyedited version of an article accepted for publication in Journal of Instrumentation. IOP Publishing Ltd is not responsible for any errors or omissions in this version of the manuscript or any version derived from it. The Version of Record is available online at \url{http://dx.doi.org/10.1088/1748-0221/11/11/P11005}}}


\begin{document}
\maketitle
\flushbottom

\section{Introduction}\label{sec:intro}

Neutron spectrometry using Bonner Spheres was introduced by Bramblett et al.~\cite{bib:bramblett}. Bonner Spheres can measure neutron energy spectra in a wide energy range from thermal up to at least 20 MeV \citep{bib:kralik}. The energy range can be extended above 1 GeV if some high atomic number materials, such as lead, copper, iron, tungsten, etc., is included into the spheres \citep{bib:hsu, bib:birattari, bib:birattari-linus, bib:wiegel}. Neutron detection using Bonner Spheres also has the advantages of high detection efficiency, simple electronics, simple operation, isotropic angular response, and excellent photon discrimination \citep{bib:alevra, bib:thomas}. On the other hand, it has the drawback of poor energy resolution. In addition, the spectral unfolding process, the process which determines the energy spectrum from detector response, is generally complex \citep{bib:alevra, bib:thomas}. Bonner Sphere neutron spectrometers can be used in, for example, workplaces around nuclear reactors, high-energy accelerators, fabrication plants of radioactive sources, and flight altitude where the knowledge of neutron energy spectrum or dose is important.

In this paper, we present our work of developing a Bonner Sphere neutron spectrometer and a spectral unfolding code. The hardware configuration of the detector and the data acquisition system will be described in sections \ref{sec:config} and \ref{sec:daq}, respectively. The determination of the detector response and the detection background will be discussed in sections \ref{sec:response} and \ref{sec:background}, respectively. Section \ref{sec:unfold} outlines the spectral unfolding procedure. The result of the development will be discussed in section \ref{sec:results}, followed by a conclusion in section \ref{sec:conclusion}.

\section{Detector configuration}\label{sec:config}

The Bonner Sphere neutron spectrometer (BSS) we developed consists of a $^{3}$He thermal-neutron detector, a set of eight polyethylene spherical shells for neutron moderation and two optional lead shells. The BSS utilizes a Berthold LB 6411 neutron-dose-rate meter \citep{bib:burgkhardt, bib:klett, bib:berthold} as the thermal-neutron detector. The active detector is a cylindrical $^{3}$He proportional counter tube. The diameter and the active length of the counter tube were designed to be both 4 cm, resulting in a better isotropic response \citep{bib:klett}. The counter tube was made of stainless steel and filled with $^{3}$He and methane at partial pressures of 3.5 bar and 1 bar, respectively \citep{bib:klett}.

\begin{figure}
	\centering
		\includegraphics[width=4.0in]{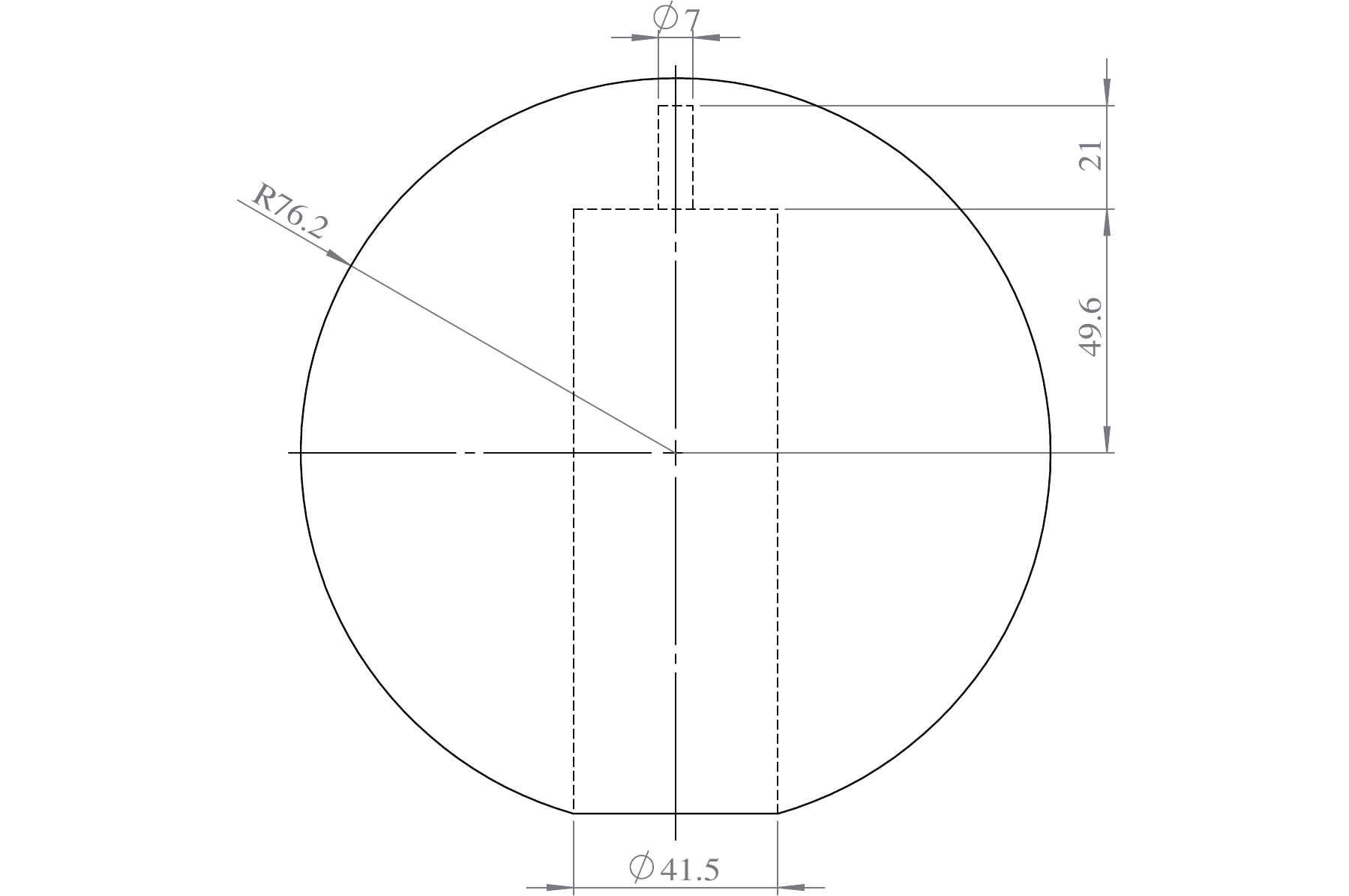}
	\caption{Mechanical drawing of the Bonner Sphere with diameter of 6" (15.24 cm). The central hollow cylinder is for inserting the thermal neutron detector of Berthold LB 6411.}
	\label{fig:bss-drawing}
\end{figure}

\begin{figure}
	\centering
		\includegraphics[width=4.0in]{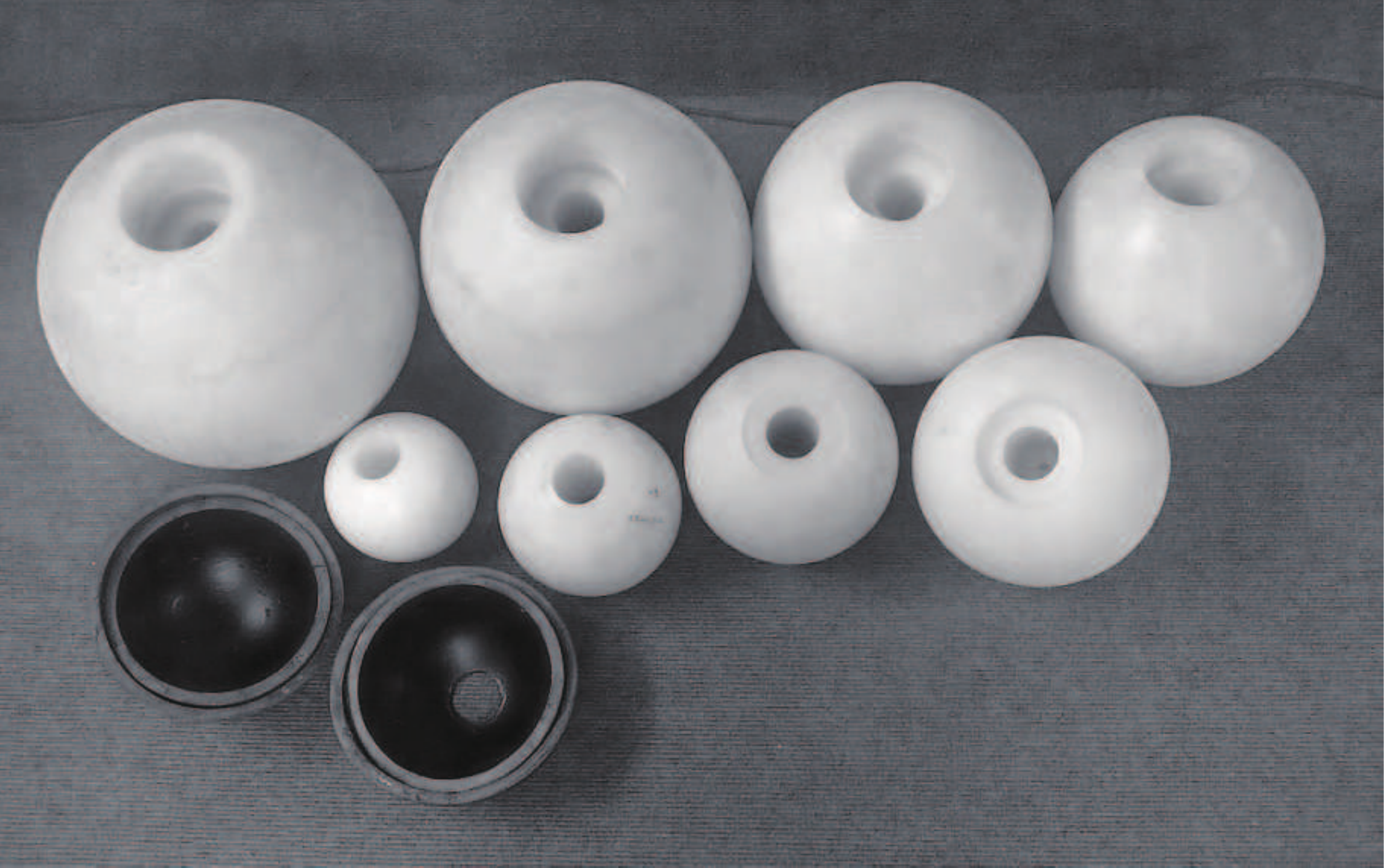}
	\caption{Eight Bonner Spheres with diameters from 5" (12.7 cm) to 12" (30.48 cm), and four 1-cm-thick lead half shells (bottom left). The lead half shells can be combined to form either a 1-cm or a 2-cm-thick lead shell surrounding the 6" Bonner Sphere.}
	\label{fig:bss-spheres}
\end{figure}

The moderating spheres of the BSS were made of polyethylene with mass density of 0.96 g cm$^{-3}$. The outer diameters of the eight spheres are 5" (12.7 cm), 6" (15.24 cm), 7" (17.78 cm), 8" (20.32 cm), 9" (22.86 cm), 10" (25.40 cm), 11" (27.94 cm), and 12" (30.48 cm), respectively.\footnote{The convention of labeling Bonner Spheres by their diameters in inches has developed over the years and it is the convention adopted here.} Figure \ref{fig:bss-drawing} shows a mechanical drawing of the 6"-diameter Bonner Sphere. Each sphere has an identical cylindrical hollow region for inserting the counter tube. The active volume of the counter tube is located at the center of each sphere, regardless of the diameter of the sphere. In order to increase the detector responses to neutrons above 10 MeV, spherical lead shells of either 1-cm or 2-cm thickness can be added to enclose the 6" sphere. Therefore, ten detector configurations with different neutron energy responses can be achieved. Figure \ref{fig:bss-spheres} shows a picture of the Bonner Spheres.

\section{Data acquisition}\label{sec:daq}

The Berthold LB 6411 neutron-dose-rate meter has an integrated high-voltage supply and front-end electronics such as preamplifier and discriminator for signal processing \citep{bib:burgkhardt, bib:klett}. It is connected to a microprocessor-controlled portable data logger Berthold UMo LB 123. With the original LB 6411 counter, the data logger can give the neutron dose rate as well as the detection count rate. When the LB 6411 was used as a thermal neutron detector of the BSS, the original neutron moderator sphere was replaced by custom-made moderator spheres of different sizes, thus the neutron dose rate reported by the data logger was no longer valid. The requirement of the BSS data acquisition (DAQ) system is to record the neutron detection count rates when the spectrometer is equipped with different sizes of moderator.

\begin{figure}
	\centering
		\includegraphics[width=4.0in]{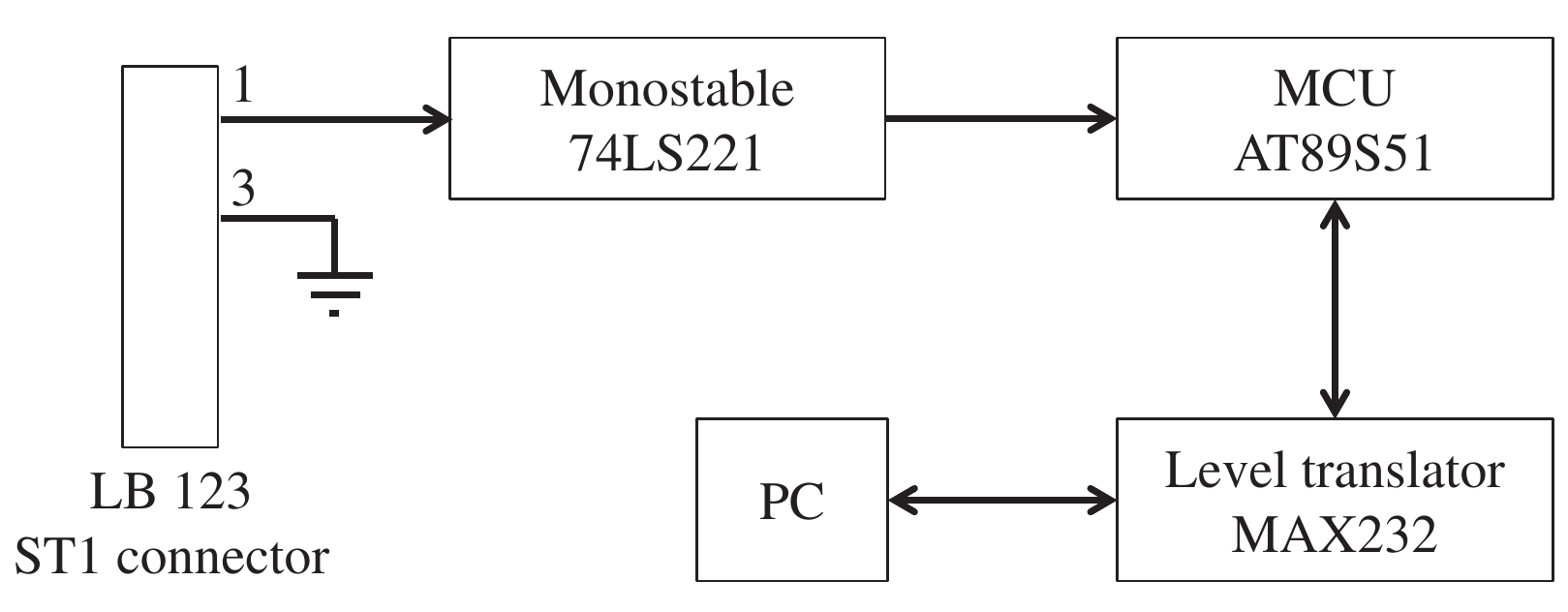}
	\caption{Schematic diagram of the computerized DAQ system. The TTL signal is extracted from pin-1 of the ST1 connector inside the LB 123. The signal levels are referenced to pin-3 of the ST1 connector.}
	\label{fig:bss-daq}
\end{figure}

The battery-driven data logger LB 123 has 250 memory locations to store the average detection count rate over the measurement period. The maximum time for each data acquisition run is limited to 99,999 s. In order to extend the capability of the DAQ system, a computerized DAQ system was added on top of the LB 123. A simplified schematic diagram of the computerized DAQ system is shown in figure \ref{fig:bss-daq}. A detection in the form of a TTL signal is extracted from the LB 123 and passed to a microcontroller unit (MCU), with a monostable in between to widen the TTL pulse width to 3 $\mu$s. The MCU has an internal counter to count the number of TTL signals. The MCU is controlled by a computer via the RS232 interface and is powered by the computer's USB port. A DAQ software written in the C programming language is used to read the accumulated count from the counter every second and to provide basic run control. If the count read from the counter is greater than the previous cached value, it will be written to an ASCII-formatted file together with a time stamp. The data files show the time-ordered values of the accumulated count, so that the count rate in various time intervals can be calculated. Every data file has a header which records the start time, measurement location, BSS configuration, and parameters such as preset time limit and preset count limit. The LB 123 was also modified to be able to obtain power from the computer's USB port. The modified DAQ system is able to run over an extended period of time beyond the default 99,999 s.

\section{Detector responses}\label{sec:response}

The response of a Bonner Sphere is defined as the ratio of the expected Bonner Sphere reading to the neutron fluence at the point where the center of the sphere is placed in the absence of the sphere \citep{bib:alevra}. Assuming both an isotropic neutron field and BSS response, the response $R_{d}(E)$ (in cm$^{2}$) of a Bonner Sphere to neutrons of energy $E$ is given by
\begin{equation}
	R_{d}(E) = \frac{B_{d}}{\phi(E)} \, ,
\end{equation}
where $B_{d}$ is the reading of the sphere (in counts), $\phi(E)$ is the incident neutron fluence (in neutrons cm$^{-2}$), and the subscript $d$ is an index identifying the Bonner Sphere. If the Bonner Sphere is exposed to a neutron field with spectral fluence $\phi_{E}(E)$, the reading of each sphere can be obtained by integrating (folding) its fluence response with the spectral fluence, that is, 
\begin{equation}
	B_{d} = \int_{0}^{\infty} \! R_{d}(E) \, \phi_{E}(E) \, \textnormal{d}E \, .
	\label{eq:bss-fredholm}
\end{equation}
When processing the BSS data, the integral in eq.~\eqref{eq:bss-fredholm} is approximated by a quadrature sum: 
\begin{align}
	B_{d} &\cong \sum_{g=1}^{n_{E}} R_{d}(E_{g}) \, \phi_{E}(E_{g}) \, \Delta E_{g} \\
	&= \sum_{g=1}^{n_{E}} R_{d,g} \, \phi_{g} \, ,
	\label{eq:bss-quadrature-sum}
\end{align}
where $\phi_{g} = \phi_{E}(E_{g}) \, \Delta E_{g}$ is the total neutron fluence in the $g^{\textnormal{th}}$ energy group of width $\Delta E_{g}$, $R_{d,g} = R_{d}(E_{g})$ is the fluence response of Bonner Sphere $d$ to neutrons in energy group $g$, and $n_{E}$ is the number of energy groups.

\subsection{Calculation of the response matrix}\label{sec:response-calculate}

\begin{figure*}
	\centering
		\includegraphics[width=\linewidth]{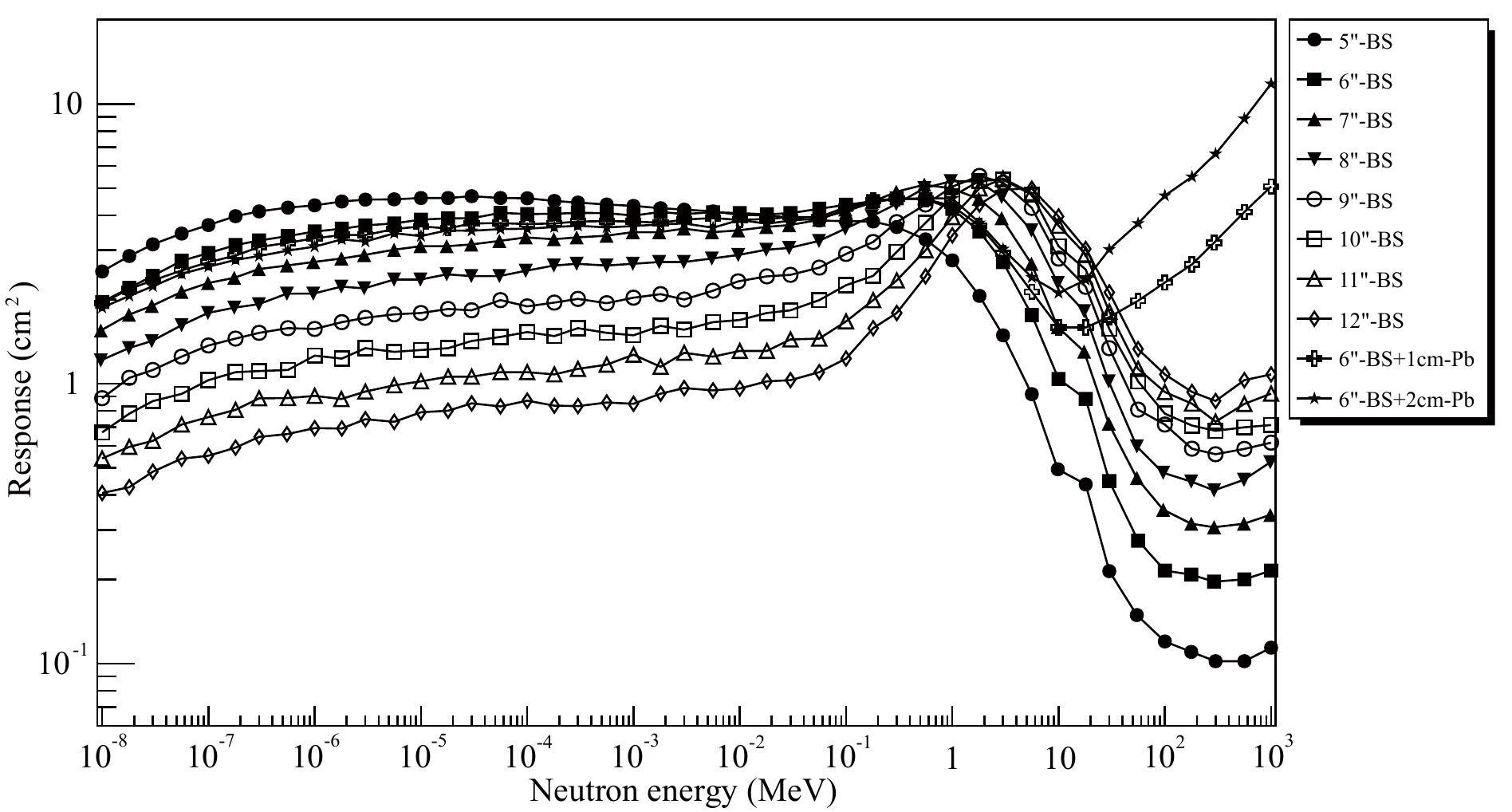}
	\caption{Neutron fluence response matrix of the Bonner Sphere neutron spectrometer calculated with GEANT4.}
	\label{fig:bss-response}
\end{figure*}

The response functions of the BSS were calculated with GEANT4 simulation code \citep{bib:agostinelli} version 9.4 (patch-02). The neutron data library G4NDL 3.14 with thermal-neutron cross sections was used. In the simulation, detailed geometry and composition of the Bonner Spheres and the $^{3}$He proportional counter tube were implemented, and fluence responses were calculated for the eight Bonner Spheres and the two configurations with the two lead shells. The calculations were performed for 45 logarithmic equidistant energy points between 10 meV and 1 GeV. For each Bonner Sphere configuration and each energy point, one million monoenergetic neutrons were generated uniformly on a disc which had the same diameter as the sphere being considered. The neutrons were transported in the same direction towards the sphere and the number of $^{3}$He$(n,p)$T reactions that occurred in the active volume of the counter tube was recorded. The fluence response was the recorded number divided by the neutron fluence of the plane source. The calculated fluence response matrix of the BSS is shown in figure \ref{fig:bss-response}. For the Bonner Spheres without the lead shells, the peak response energy increases with the size of the sphere. Addition of the lead shells improves the responses to neutron energies higher than 20 MeV due to the extra $(n,xn)$ reactions in lead.

\subsection{Verification of the response matrix}\label{sec:response-verify}

A good knowledge of the fluence response matrix of the BSS is crucial for obtaining reliable spectrometric results \citep{bib:alevra-bss-influence}. In order to validate the GEANT4 simulation, a series of measurements with an $^{241}$Am-Be$(\alpha,n)$ neutron source was performed. The source produces neutrons predominantly through the following reactions: 
\begin{equation}
	^{9}\textnormal{Be} + \alpha \to ^{13}\textnormal{C}^{*} \to ^{12}\textnormal{C} + n + \gamma\textnormal{(4.4 MeV)} \, ,
	\label{eq:ambe-reaction-w-gamma}
\end{equation}
\begin{equation}
	^{9}\textnormal{Be}(\alpha,\alpha')^{9}\textnormal{Be} \to ^{8}\textnormal{Be} + n \, .
	\label{eq:ambe-reaction-wo-gamma}
\end{equation}
The ratio of the 4.4-MeV gamma-ray intensity to the neutron intensity is a characteristic of $^{241}$Am-Be sources. The gamma emission rate of the $^{241}$Am-Be source was measured using a calibrated high-purity germanium (HPGe) detector. The source was placed next to the aluminum shell of the HPGe detector and measured for a live time of $\tau = 316268$ s. The absolute efficiencies of detecting the full-energy peak, the single-escape peak, and the double-escape peak of the 4.4-MeV gamma rays were studied by irradiating $N_{\gamma} = 1 \times 10^{7}$ gamma photons in a GEANT4 simulation with the same configuration as the experiment. A linear regression of the net measured counts of the three gamma-ray peaks versus the simulated counts gave a proportionality constant of $1.19 \pm 0.08$. Thus, the emission rate of 4.4-MeV gamma-rays of the $^{241}$Am-Be source was determined to be $(1.19 \pm 0.08) \times N_{\gamma} / \tau = (37.6 \pm 2.7)$ s$^{-1}$. Liu et al.~\cite{bib:liu-ambe} measured and gave a review of the 4.4-MeV gamma-to-neutron intensity ratio $R$ of $^{241}$Am-Be sources. We adopted their recommended value of $R = 0.575 \pm 0.028$. Therefore, the neutron emission rate of the $^{241}$Am-Be source was determined to be $(37.6 \pm 2.7)/R = (65.3 \pm 5.6)$ s$^{-1}$.

\begin{figure}
	\centering
		\includegraphics[width=4.0in]{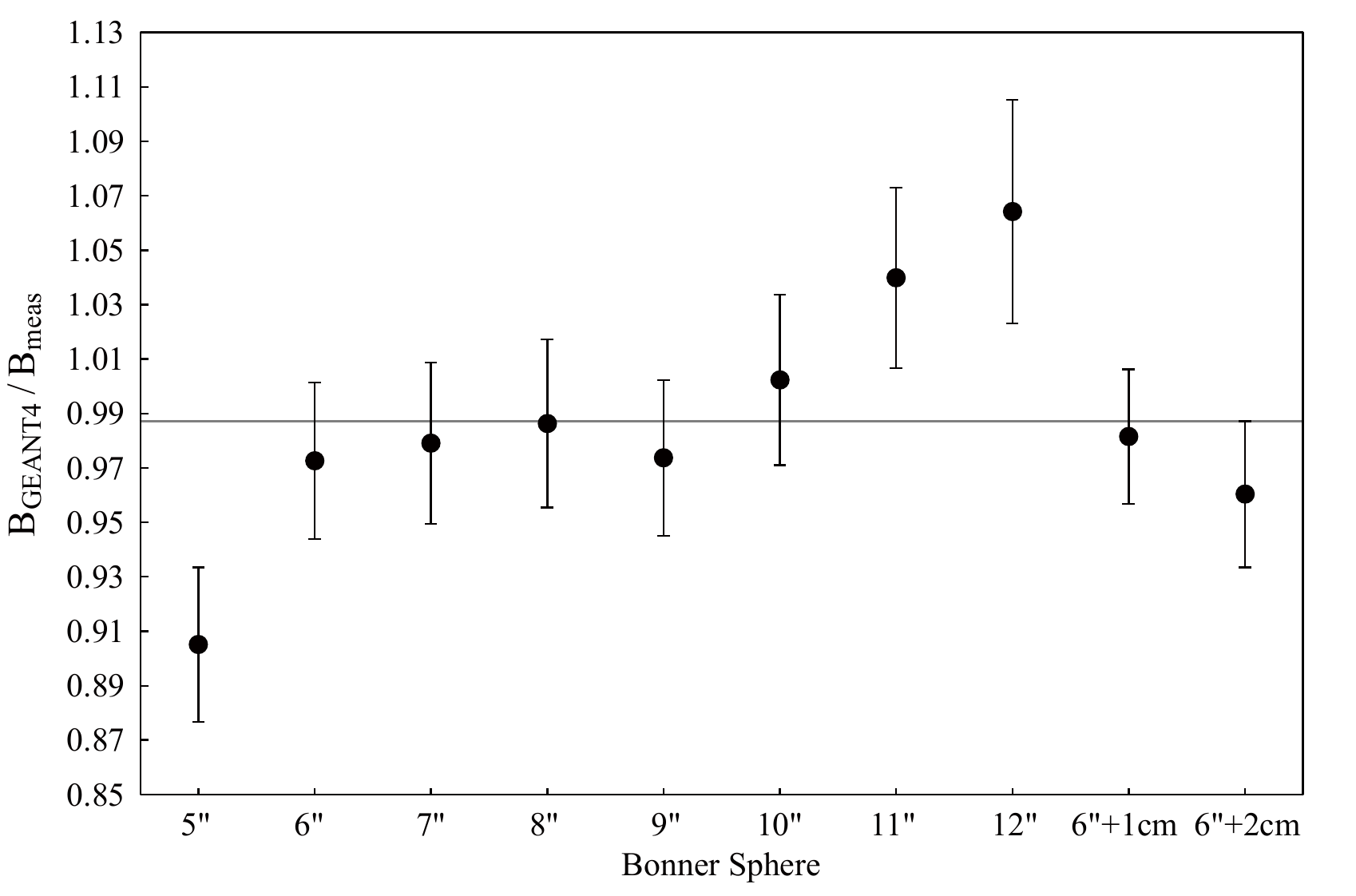}
	\caption{Ratios of simulated to measured count rates from table \ref{tab:bss-ambe-compare}. The errors were propagated from the statistical uncertainties of the count rates. The grey line shows the average ratio of 0.986.}
	\label{fig:bss-ambe-ratio}
\end{figure}

The $^{241}$Am-Be source was attached to the surface of each Bonner Sphere in turn and the resulting measured count rates $B_{\textnormal{meas}}$ are tabulated in table \ref{tab:bss-ambe-compare}. The same set up with the $^{241}$Am-Be source was implemented in a GEANT4 simulation. In the simulation, one million neutrons were generated isotropically from the source following the $^{241}$Am-Be neutron energy spectrum recommended by the International Organization for Standardization (ISO) \citep{bib:iso-ambe}. The simulated counts were scaled to the equivalent count rates $B_{\textnormal{GEANT4}}$ using the determined neutron emission rate of the $^{241}$Am-Be source, and the results are also listed in table \ref{tab:bss-ambe-compare}. The ratios $B_{\textnormal{GEANT4}} / B_{\textnormal{meas}}$ calculated from the count rates are shown in figure \ref{fig:bss-ambe-ratio}. On average, the simulated count rates were about $1.4\%$ lower than the measured count rates, but this was within the $\pm 8.6\%$ uncertainty of the neutron emission rate of the $^{241}$Am-Be source. The variability (one standard deviation) of the ratios was $\pm 4.1\%$, suggesting a high level of accuracy in the GEANT4 simulation. The same level of accuracy could possibly be assigned to the response matrix simulation.

\begin{table}
	\small
	\centering
	\caption{Measured ($B_{\textnormal{meas}}$) and simulated ($B_{\textnormal{GEANT4}}$) count rates of each Bonner Sphere when the $^{241}$Am-Be neutron source was attached to the surface. Errors were statistical uncertainties assuming Poisson distribution of counts.}
		\begin{tabular}{ l l l }
			\toprule
			Bonner Sphere & $B_{\textnormal{meas}}$ ($10^{-1}$ s$^{-1}$) & $B_{\textnormal{GEANT4}}$ ($10^{-1}$ s$^{-1}$) \\
			\midrule
			5''-BS & $2.95 \pm 0.08$ & $2.67 \pm 0.04$ \\
			6''-BS & $3.12 \pm 0.08$ & $3.03 \pm 0.04$ \\
			7''-BS & $3.03 \pm 0.08$ & $2.97 \pm 0.04$ \\
			8''-BS & $2.60 \pm 0.07$ & $2.56 \pm 0.04$ \\
			9''-BS & $2.13 \pm 0.05$ & $2.07 \pm 0.04$ \\
			10''-BS & $1.66 \pm 0.04$ & $1.66 \pm 0.03$ \\
			11''-BS & $1.30 \pm 0.03$ & $1.35 \pm 0.03$ \\
			12''-BS & $1.01 \pm 0.03$ & $1.07 \pm 0.03$ \\
			6''-BS+1cm-Pb & $2.58 \pm 0.05$ & $2.53 \pm 0.04$ \\
			6''-BS+2cm-Pb & $2.28 \pm 0.05$ & $2.19 \pm 0.04$ \\
			\bottomrule
		\end{tabular}
	\label{tab:bss-ambe-compare}
\end{table}

\section{Neutron detection backgrounds}\label{sec:background}

\subsection{Detector background}\label{sec:detector-bkg}

\begin{figure}
	\centering
		\includegraphics[width=4.0in]{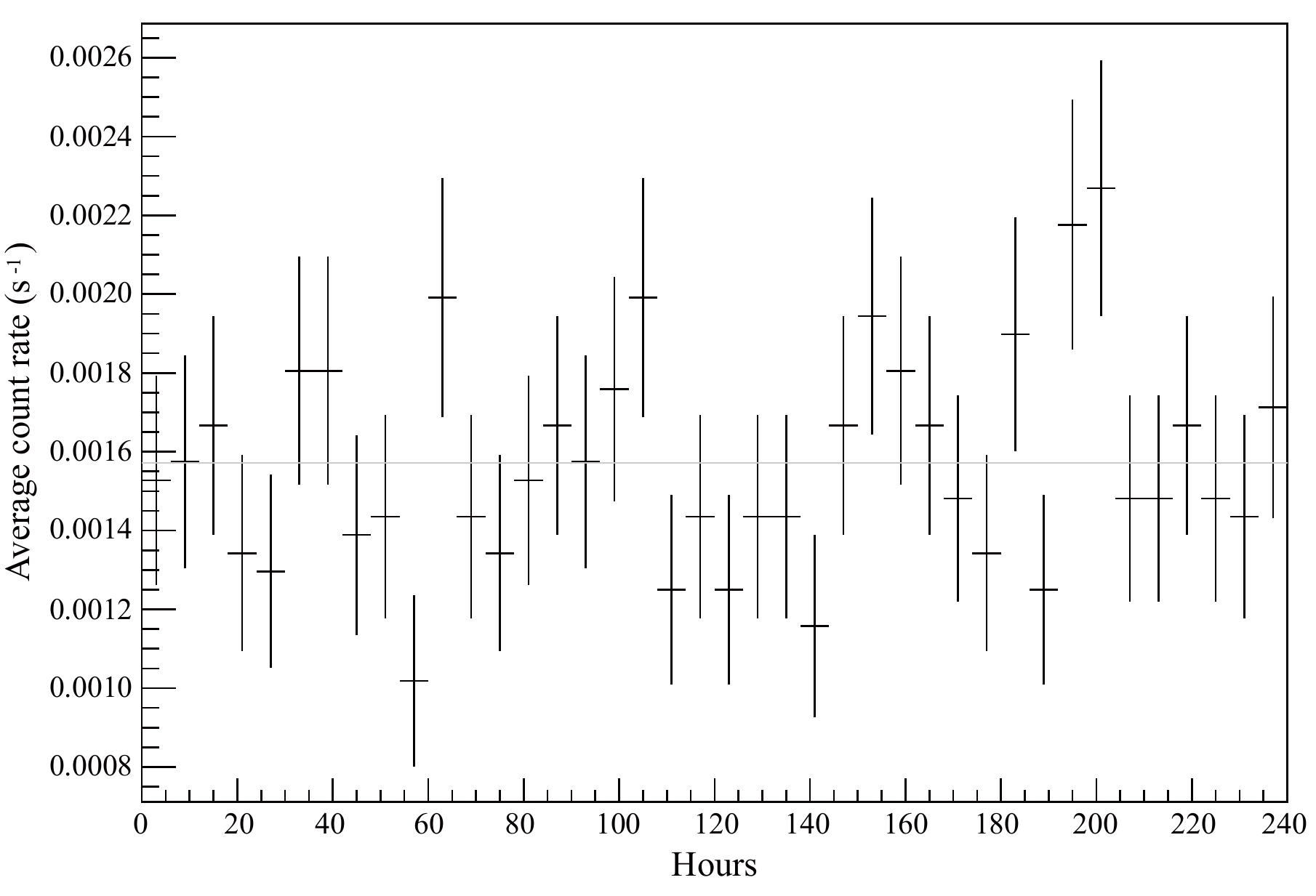}
	\caption{Detector background rates of the BSS as a function of time. The number of signals was binned by every 6 hours. The grey line shows the average rate of $(1.57 \pm 0.04) \times 10^{-3}$ s$^{-1}$.}
	\label{fig:bss-bkg-rate}
\end{figure}

The detector background rate of the $^{3}$He proportional counter was measured by shielding the counter tube from ambient neutrons. The counter tube was submerged in borax (molecular formula: Na$_{2}$B$_{4}$O$_{7} \cdot 10$H$_{2}$O) powders with minimum thickness of 30 cm in all directions. Simulations showed that the amount of borax used could effectively reduce the detection of ambient neutrons by almost four orders of magnitude. The measurement set up was put in an underground environment with 611 meters water equivalent of overburden to further reduce the influences of neutrons induced by cosmic-ray muons. The time series of the detector background rates is shown in figure \ref{fig:bss-bkg-rate}. No obvious trends of systematic fluctuations could be observed. The average detector background rate of the BSS was $(1.57 \pm 0.04) \times 10^{-3}$ s$^{-1}$.

\subsection{Gamma-ray discrimination}\label{sec:gamma-bkg}

The gamma-ray discrimination factor of the thermal-neutron detector was tested with a $^{60}$Co gamma-ray source, which emits two gamma rays at energies of 1.17 and 1.33 MeV, respectively. The $^{60}$Co source was placed at 2 cm away from the bare counter tube, resulted in a rate of $2.7 \times 10^{5}$ gamma rays penetrating the $^{3}$He active volume per second. With the $^{60}$Co source in place, the count rate was increased by $(3.9 \pm 2.2) \times 10^{-3}$ s$^{-1}$ in one hour of measurement. Therefore, the lower limit of the gamma-ray discrimination factor in a $^{60}$Co gamma-ray field was determined to be $2 \times 10^{7}$.

\section{Unfolding of neutron spectra}\label{sec:unfold}

\subsection{The unfolding problem}\label{sec:unfold-problem}

If a neutron field is measured with a set of $n_{D}$ Bonner Spheres, we obtain a set of $n_{D}$ readings. Each reading $B_{d}$ originates from a linear combination of neutron fluence and detector response like that in eq.~\eqref{eq:bss-quadrature-sum}. The set of these $n_{D}$ equations can be written in a matrix form
\begin{equation}
	\mathbf{B} = \mathbf{R} \, \mathbf{\Phi} \, ,
	\label{eq:bss-matrix-eq}
\end{equation}
where $\mathbf{B}$ is the reading vector with $n_{D}$ components $B_{d} \ (d = 1, \ 2, \ ..., \ n_{D})$, $\mathbf{\Phi}$ is a vector which contains the spectral fluence information with $n_{E}$ energy groups $\phi_{g} \ (g = 1, \ 2, \ ..., \ n_{E})$, and $\mathbf{R}$ is the $n_{D} \times n_{E}$ rectangular fluence response matrix. The matrix $\mathbf{R}$ can be seen as an operator which transforms the information from fluence space to reading space during the measurement process. We are interested in determining the fluence $\phi_{g}$ for all energy groups given a number of measurement readings $B_{d}$ and the fluence response matrix $\mathbf{R}$ of the BSS. Since the number of energy groups is normally greater then the number of measurement readings ($n_{E} > n_{D}$) and the response functions of the BSS are considerably overlapped, the matrix problem is underdetermined leading to an infinite number of possible $\mathbf{\Phi}$ which satisfies eq.~\eqref{eq:bss-matrix-eq}. Therefore, the goal of the unfolding process is to find a single solution $\mathbf{\Phi}$ which closely approximates the actual neutron energy spectrum.

Several ways by which the unfolding problem may be solved and a number of representative unfolding codes are described in ref.~\citep{bib:matzke}. Many of the unfolding codes require an input guess spectrum which is iteratively adjusted until a solution spectrum is obtained \citep{bib:alevra}.

\subsection{Search space}\label{sec:search-space}

The search space defines the range of possible neutron fluence in each energy group. Obviously, the lower fluence bound $\mathbf{\Phi}_{\mathnormal{min}}$ is \textbf{0} and the upper bound $\mathbf{\Phi}_{\mathnormal{max}}$ is related to the measured readings of the BSS. The upper bound $\phi_{max,g}$ for any energy group $g$ can be constructed from the measured reading of all Bonner Spheres by assuming that all of the fluence is concentrated in that particular energy group, that is, 
\begin{equation}
	\phi_{max,g} = \min\{ \frac{B_{d}}{R_{d,g}} \ | \ \forall d \in [1,n_{D}] \} \, .
	\label{eq:bss-ss}
\end{equation}
Minimum values are taken in eq.~\eqref{eq:bss-ss} for a tighter search space. Although the above method is rigorous, it is too conservative for most part of the spectrum unless it is dominated by a handful of monoenergetic energy peaks. The search space can be reduced by assuming that during the construction process all of the fluence is concentrated but spread out over a handful of energy groups, instead of all concentrated in a single energy group. The reduced upper bound $\mathbf{\Phi}_{\mathnormal{max}}^{\mathnormal{G}}$ is therefore
\begin{equation}
	\phi_{max,g}^{G} = \min\{ \frac{B_{d}}{\sum_{m=E_{L}(g)}^{E_{H}(g)} R_{d,m}} \ | \ \forall d \in [1,n_{D}] \} \, ,
	\label{eq:bss-ss-g}
\end{equation}
with the lower ($E_{L}$) and upper ($E_{H}$) bounds of the range of energy groups following
\begin{align}
	E_{L}(g) &= \max\{ g-\left\lfloor G/2 \right\rfloor, 1 \} \, , \\
	E_{H}(g) &= \min\{ g+\left\lceil G/2 \right\rceil-1, n_{E} \} \, ,
\end{align}
where $G \in \mathbb{N}$ is the number of energy groups where the fluence is spread. Equation \eqref{eq:bss-ss-g} reduces to eq.~\eqref{eq:bss-ss} when $G = 1$. The choice of $G$ depends on the neutron energy spectrum in question. For example, large $G$ values may greatly reduce the search space, but may result in the clipping of some prominent energy peaks. A suitable choice of $G$ is discussed in section \ref{sec:result-param}.

\subsection{Genetic algorithm spectral unfolding}\label{sec:nsuga}

Genetic algorithm was introduced by Holland \cite{bib:holland} to mimic the process of natural evolution. It generates solutions to optimization problems using techniques which are inspired by natural evolution, such as inheritance, selection, mutation, and crossover. Each candidate solution is referred to as an ``individual'' and all candidate solutions form a ``population''. Individuals are encoded in strings of numbers which represent the ``chromosomes''. The ``worth'' of each individual is characterized by its similarity with the expected solution and is expressed by a numerical value known as ``fitness''. In each generation, multiple individuals are selected from the current population based on their fitness, and modified by the means of mutation and recombination of chromosomes to form a new population. The new population is used in the next iteration. The evolution process continues until a satisfactory fitness has been reached by any individual, or the number of generations has reached a prescribed maximum.

Freeman et al.~\cite{bib:freeman} applied genetic algorithms to the Bonner Sphere neutron spectrum unfolding problem and developed the UMRGA code. The code was shown to perform well in the absence of \textit{a priori} information such as an input guess spectrum. A multi-seed averaging technique could be implemented to improve the spectral quality by averaging solutions with different random seeds. It also had a monoenergetic peak reward technique to encourage spectral peaks at particular energies assuming that the energy levels of the peaks were known in advance. Mukherjee \cite{bib:mukherjee-bondi-97, bib:mukherjee-hi-res} utilized a commercial genetic-algorithm engine running on Microsoft Excel to develop the BONDI-97 unfolding code. The code was able to reproduce the integrated neutron fluence rate and the dose-equivalent rate of an $^{241}$Am-Be source. Recently, Wang et al.~\cite{bib:wang} developed an unfolding code using genetic algorithms with a pseudo-parallel strategy to prevent premature convergence. They tested two different definitions of fitness and showed that the choice of fitness functions had a crucial effect in the unfolded spectra.

\begin{figure*}
	\centering
		\includegraphics[width=\linewidth]{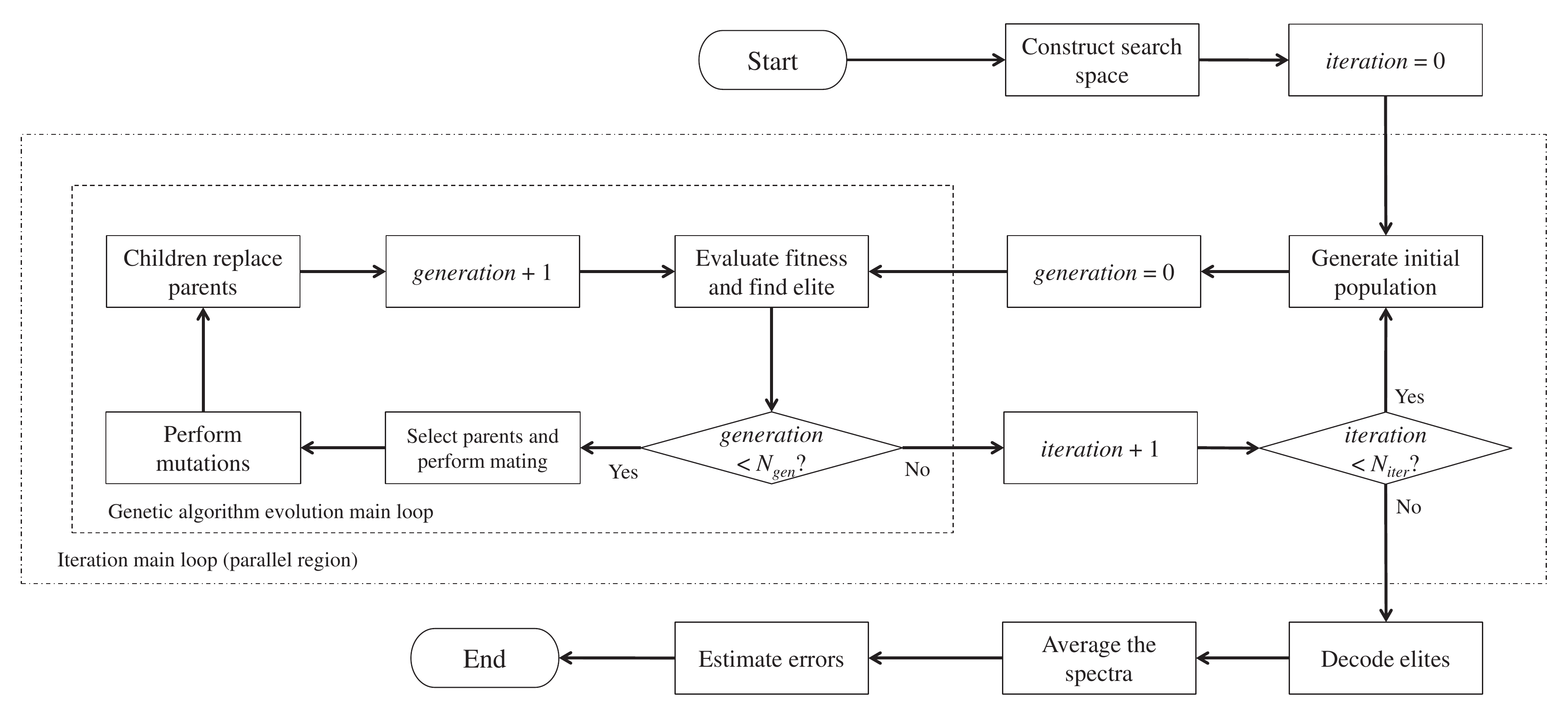}
	\caption{Flow chart of the spectral unfolding procedures of NSUGA.}
	\label{fig:nsuga-flow}
\end{figure*}

In conjunction of the development of the BSS, a spectral unfolding code NSUGA (Neutron Spectrum Unfolding by Genetic Algorithm) was developed based on the ideas proposed by Freeman et al.~\cite{bib:freeman}. NSUGA, written in the C++ programming language, is implemented with a multi-seed averaging technique and a peak reward technique similar to UMRGA, but the peak reward is extended to include energy peaks of different spectral widths. NSUGA uses a fitness definition based on Poisson's statistics, which is naturally associated with the neutron and background detection processes. Furthermore, it employs a Monte Carlo technique to estimate the uncertainty in each energy group in the unfolded spectra. Taking the advantages of modern multi-thread and multi-core computers, NSUGA uses parallel processing to reduce the unfolding time. The spectral unfolding procedures of NSUGA is illustrated in figure \ref{fig:nsuga-flow}. The following subsections describe the key concepts of the building blocks of NSUGA.

\subsubsection{Encoding and decoding of solutions}\label{sec:encode-decode}

Since the fluences in a neutron energy spectrum can span a few orders of magnitude, the fluences are digitized to reduce the number of possible values while retaining the dynamic range. The search space of each energy group is linearly partitioned into $N_{part} = 2^{8} = 256$ discrete values between $\phi_{min,g}$ and $\phi_{max,g}^{G}$. The $i^{\textnormal{th}}$ candidate fluence value, $\phi_{i,g}$, in energy group $g$ can be calculated as
\begin{equation}
	\phi_{i,g} = \left( \frac{i-1}{N_{part}-1} \right) \left( \phi_{max,g}^{G} - \phi_{min,g} \right) + \phi_{min,g} \, ,
	\label{eq:nsuga-phi-i}
\end{equation}
for $i \in [1,N_{part}]$. Therefore, each candidate solution in NSUGA can be defined by $n_{E}$ integers and the chromosome $\mathbf{K}$ is defined by an array of 8-bit unsigned integers as: 
\begin{equation}
	\mathbf{K} = \{ \mathnormal{K_{g}} \ | \ \forall g \in [\textnormal{1},n_{E}] \} \, ,
	\label{eq:nsuga-chrome}
\end{equation}
where $K_{g} \in [0,N_{part}-1]$ determines the fluence of the $g^{\textnormal{th}}$ energy group. In the case of $\mathbf{\Phi}_{\mathnormal{min}} = \textbf{0}$, eq.~\eqref{eq:nsuga-phi-i} can be reduced to
\begin{equation}
	\phi_{g} = \left( \frac{K_{g}}{N_{part}-1} \right) \phi_{max,g}^{G} \, .
	\label{eq:nsuga-phi-i-reduce}
\end{equation}

\subsubsection{Initial population}\label{sec:population}

The chromosomes of the initial population are filled with random numbers ranged from 0 to $N_{part}-1$. A fundamental issue in genetic algorithms is to decide an appropriate population size $N_{pop}$. Carroll \cite{bib:carroll-psize, bib:carroll-pmute} recommended the following population sizing equation based on the work presented by Goldberg et al.~\cite{bib:goldberg}: 
\begin{equation}
	N_{pop} = O\left[ n_{E} \cdot \chi^{k} \right] \, ,
\end{equation}
where $\chi$ is the cardinality of chromosomes (for binary coding, $\chi = 2$) and $k$ is the number of bits in each parameter. In our case where binary coding is chosen, $\chi^{k} = N_{part}$, thus
\begin{equation}
	N_{pop} = O\left[ n_{E} \cdot N_{part} \right] \, .
	\label{eq:nsuga-psize}
\end{equation}
With the number of energy groups $n_{E} = 45$ (as the same one used in the calculation of the response matrix) and taking $N_{part} = 2^{8} = 256$, the population size $N_{pop} = 45 \times 256 = 11520$ is used as the default value in NSUGA.

\subsubsection{Fitness evaluation}\label{sec:fitness}

Genetic algorithms search the optimum solution based on Darwin's theory of natural selection. An individual which performs well in its environment has a higher probability to survive and pass its genetic information to the next generation. In NSUGA, every individual in the population is evaluated for its fitness at the beginning of each generation. For our purpose of Bonner Sphere spectral unfolding, fitness is a measure of how well a candidate solution can reproduce the measured readings of the Bonner Spheres. To determine the fitness of an individual, the chromosome is first decoded into fluence values by eq.~\eqref{eq:nsuga-phi-i-reduce}, then the decoded spectral fluence is folded with the fluence response matrix to form a set of calculated sphere readings $B_{calc,d}$, where
\begin{equation}
	B_{calc,d} = \sum_{g=1}^{n_{E}} R_{d,g} \, \left( \frac{K_{g}}{N_{part}-1} \right) \phi_{max,g}^{G} \, .
\end{equation}
The probability of observing $B_{d}$ counts while expecting $B_{calc,d}$ counts follows a Poisson distribution: 
\begin{equation}
	p(B_{d};B_{calc,d}) = \frac{(B_{calc,d})^{B_{d}} \cdot e^{-B_{calc,d}}}{B_{d}!} \, .
\end{equation}
If we include the contribution of detector backgrounds during measurements, the joint likelihood of observing a set of BSS counts $\mathbf{B}$ while expecting $\mathbf{B}_{\mathnormal{calc}}$ can be written as: 
\begin{equation}
	L = \prod_{d=1}^{n_{D}} \left[ \sum_{c=1}^{\infty} p(B_{d};B_{calc,d}+c\cdot\frac{T_{d}}{T_{BG}}) \cdot p(B_{BG};c) \right] \, ,
\end{equation}
where $B_{BG}$ is the measured detector background count, and $T_{d}$ ($T_{BG}$) is the measurement time of Bonner Sphere $d$ (detector background). The fitness function $f$ is defined as the average log-likelihood: 
\begin{equation}
	f = 100 + \frac{1}{n_{D}} \log L \, .
	\label{eq:nsuga-fitness}
\end{equation}
The fitness increases with increasing agreement between the measured and the calculated readings. The value of 100 is added to eliminate negative $f$ values and to scale $f$ so that perfect agreement between measurements and expectations will result in a fitness of 100. The goal of the genetic algorithm is to maximize $f$ in a reasonable amount of time.

\subsubsection{Evolution}\label{sec:evolution}

Apart from fitness evaluation, the evolution cycle also includes processes such as parent selection, mating, mutation, and a replacement operation.

Selections of parents are biased such that a healthier (higher fitness value) candidate has a higher chance to be a parent. A particular technique known as binary tournament selection \citep{bib:goldberg-selection} is used. In each tournament, a pair of individuals are selected randomly from the population, and the individual with higher fitness is selected to be a parent. The tournament is repeated to find the mate. Tournament selection has a good time complexity of $O(n)$ \citep{bib:goldberg-selection}.

Because of the tournament selection, individuals which are selected to be parents may possess some strong genetic materials. Parents mate by mixing their genetic information to form new individuals in the hope of producing a stronger candidate solution. The mixing is done using a technique known as uniform crossover. The crossover process is random and a child is produced according to $n_{E}$ randomly generated binary bits. For example, if the $g^{\textnormal{th}}$ random bit is 0, then the chromosome for energy group $g$ of the child is inherited from parent 1, otherwise it is inherited from parent 2. The processes of parent selection and reproduction are repeated until a new population of $N_{pop}$ children is formed.

The children then undergo mutation which is carried out by randomly and infrequently changing the chromosome value of one or more energy groups. Since the changes are random, a gain in fitness after the mutation is not guaranteed and may result in worse performances. However, mutation can explore unexplored region within the search space. Two different types of mutation operations are defined in NSUGA: jump mutation and creep mutation. Jump mutation is performed by randomly changing the chromosome value of a randomly selected energy group. Jump mutation can result in a dramatic change of value within the search space of that particular energy group. On the contrary, creep mutation changes the chromosome value by a relatively small amount. It either increases or decreases the chromosome value of a randomly selected energy group by one step. The probability of jump and creep mutation of each individual in each generation is controlled by two parameters, $P_{jump}$ and $P_{creep}$, respectively. The following settings as recommended by Carroll \cite{bib:carroll-pmute} and adopted by Freeman et al.~\cite{bib:freeman} are used throughout this study: 
\begin{align}
	P_{jump} &= 1/N_{pop} \, , \\
	P_{creep} &= 2/N_{pop} \, .
\end{align}

After the mutation process, the children population replaces its parent population and becomes the parent population in the next generation. In order to preserve the best candidate solution known as the elite, which has the highest fitness among the population, the elite solution is carried over to the next generation. This elitist strategy can ensure a non-decreasing fitness of the elite along the evolution timeline and guarantee a convergence of solution \citep{bib:he}.

The whole evolution cycle is repeated until $N_{gen}$ number of generations are produced. The best solution is decoded from the elite in the last generation.

\subsubsection{Incorporating \textit{a priori} information}\label{sec:a-priori}

The underdetermined nature of the unfolding problems leads to an infinite number of possible solutions. Although NSUGA does not require an input guess spectrum to start with, it can adopt some \textit{a priori} information of the investigated neutron field. Incorporation of \textit{a priori} knowledge of the spectrum could possibly result in a more reliable unfolded spectrum and could also provide some fine details in the spectrum such as monoenergetic peaks.

\textit{A priori} information about the neutron field can be derived from calculations or previous measurements. A drawback of BSS is its poor energy resolution \citep{bib:alevra, bib:thomas}, thus \textit{a priori} information on energy peaks can potentially help to improve the spectral quality of the unfolded spectrum. NSUGA considers \textit{a priori} energy peaks when evaluating the fitness of each candidate solution. Each \textit{a priori} energy peak is characterized by three parameters: peak position $\rho$, spectral half-width $\omega$, and minimum peak-to-valley ratio $\gamma$. One more attribute known as importance $\eta$ is added to each \textit{a priori} energy peak to quantify the additional weighting given to these peaks in the calculation of the fitness factor. NSUGA looks for a peak at energy group $g = \rho$ which satisfies the criteria
\begin{equation}
	\phi_{\rho}/\phi_{\rho-\omega} > \gamma \ \ \ \ \textnormal{and} \ \ \ \ \phi_{\rho}/\phi_{\rho+\omega} > \gamma \, ,
\end{equation}
along with the criteria of a peak in the form
\begin{equation}
	\phi_{g} > \phi_{g-1}, \forall g \in (\rho-\omega, \rho] \ \ \ \ \textnormal{and} \ \ \ \ \phi_{g} > \phi_{g+1}, \forall g \in [\rho, \rho+\omega) \, .
\end{equation}
A candidate solution which satisfies the above criteria is rewarded by giving it a rescaled fitness $f'$, where
\begin{equation}
	f' = f \times (1 + \eta/100) \, .
\end{equation}
The importance of the peak determines the amount of the reward. If a monoenergetic peak (i.e., $\omega = 1$) is defined, the unscaled search space (eq.~\eqref{eq:bss-ss}) is used for that particular energy group. Multiple \textit{a priori} peaks can be defined and rewarded separately.

\subsubsection{Error estimation}\label{sec:error}

Equation \eqref{eq:bss-matrix-eq} represents an idealized model of the unfolding problem. In real situations, an actual measured reading $\mathbf{B}_{\mathnormal{0}}$ has uncertainty and can be written as: 
\begin{equation}
	\mathbf{B}_{\mathnormal{0}} = \mathbf{B} + \mathbf{e}_{\mathnormal{0}} \, ,
\end{equation}
where $\mathbf{e}_{\mathnormal{0}}$ is a fluctuation term due to statistical and systematic uncertainties.

In BSS, two fluctuation components contribute to the uncertainty of the unfolded spectrum. The first one is the uncertainty of input quantities such as sphere readings, response functions, and any \textit{a priori} information. The second one is the ambiguity of the solution due to the underdetermined nature of the problem \citep{bib:matzke}. NSUGA employs a Monte Carlo technique \citep{bib:press-error-mc} to estimate the uncertainty of the solution due to the uncertainty of sphere readings and the ambiguity in the unfolding process.

Suppose the measurement was repeated for $N_{iter}$ times, we would get $N_{iter}$ sets of sphere readings $\mathbf{B}_{\mathnormal{i}}$ and the corresponding unfolded spectra $\mathbf{\Phi}_{\mathnormal{i}}$ ($i \in [0,N_{iter}-1]$). The sphere readings $\mathbf{B}_{\mathnormal{i}}$ would have the same probability distribution as the uncertainty $\mathbf{e}_{\mathnormal{i}}$. If the number $N_{iter}$ was large enough, the distribution of the unfolded spectra $\mathbf{\Phi}_{\mathnormal{i}}$ should represent the fluctuation in $\mathbf{e}_{\mathnormal{i}}$ as well as the ambiguity in the unfolding process.

Multiple sets of sphere readings are simulated by a Monte Carlo process. The unfolding process is repeated for $N_{iter}$ times. In each iteration, the measured readings are perturbed randomly according to their uncertainties, and the perturbed readings are unfolded. The result is a set of $N_{iter}$ synthetic sphere readings $\mathbf{B}_{\mathnormal{i}}$ and a set of the corresponding unfolded spectra $\mathbf{\Phi}_{\mathnormal{i}}$. The unfolded spectra are averaged to give the solution $\mathbf{\Phi}$. The fluence uncertainty $\epsilon_{g}$ of energy group $g$ can be estimated as
\begin{equation}
	\epsilon_{g} = \left\lbrace \left[ \frac{1}{N_{iter}} \sum_{i=0}^{N_{iter}-1} \left| \phi_{i,g} - \phi_{g} \right| \right]^{2} + \frac{1}{4} \left( \frac{\phi_{max,g}^{G}-\phi_{min,g}}{N_{part}-1} \right)^{2} \right\rbrace^{1/2} \, ,
\end{equation}
where the first term is an estimate of statistical uncertainties and fluctuations due to the ambiguity using mean absolute deviation, and the last term is the uncertainty due to the partition of search space.

\subsubsection{Parallelization and random seeds}\label{sec:random-seeds}

The $N_{iter}$ times of repetition of the unfolding tasks can be executed in parallel to reduce the processing time. The parallelization of the loop is realized by using OpenMP.\footnote{OpenMP is a portable standard for the programming of shared memory systems. A good introduction to OpenMP and parallel programming is given by Rauber and Runger \cite{bib:rauber}. More information about OpenMP and the standard definition can be found at: http://openmp.org/} The OpenMP API provides a set of complier directives which can be used to transform a sequential C++ code into parallel execution. The execution of NSUGA begins with a single thread. The initial thread reads all the required input files and performs the construction of search space. After that, it creates a certain number of new threads. The multiple iterations of the unfolding tasks are executed in parallel by all threads, including the initial thread and the new threads. The parallel region ends with an implicit synchronization of all threads to make sure that all unfolding tasks have been finished before the next step. Finally, the $N_{iter}$ unfolded spectra are averaged and the spectral uncertainties are calculated. The parallel region in the execution path is indicated by the outer dashed box in figure \ref{fig:nsuga-flow}.

The stochastic processes of NSUGA rely on the generation of random numbers, or more precisely in actual applications, pseudo-random numbers. NSUGA uses random number generators (RNG) recommended by Matsumoto and Nishimura \cite{bib:matsumoto}. Since each call to a RNG updates the internal state of the generator, separate generators have to be used in different threads to avoid conflicts and race conditions. In fact, because of the object-oriented implementation of NSUGA, every candidate solution is an object and has its own RNG. The sequence of numbers generated by a RNG is initiated by a random seed. The initial random seed $s_{i,j}$ assigned to the $j^{\textnormal{th}}$ individual in the $i^{\textnormal{th}}$ iteration is
\begin{equation}
	s_{i,j} = 17 + (N_{iter}-1)^2 \times N_{pop} + (i-1) \times N_{pop} + (j-1) \, .
\end{equation}
The value 17 is an arbitrarily chosen number. The number of iterations $N_{iter}$ is added to the calculation such that the spectra obtained with a smaller $N_{iter}$ are not necessarily a subset of the spectra obtained with a larger $N_{iter}$. This may yield a better statistical comparison between results which are obtained with different values of $N_{iter}$.

\section{Results}\label{sec:results}

\subsection{Tests of unfolding code parameters}\label{sec:result-param}

\begin{figure}
	\centering
		\includegraphics[width=4.0in]{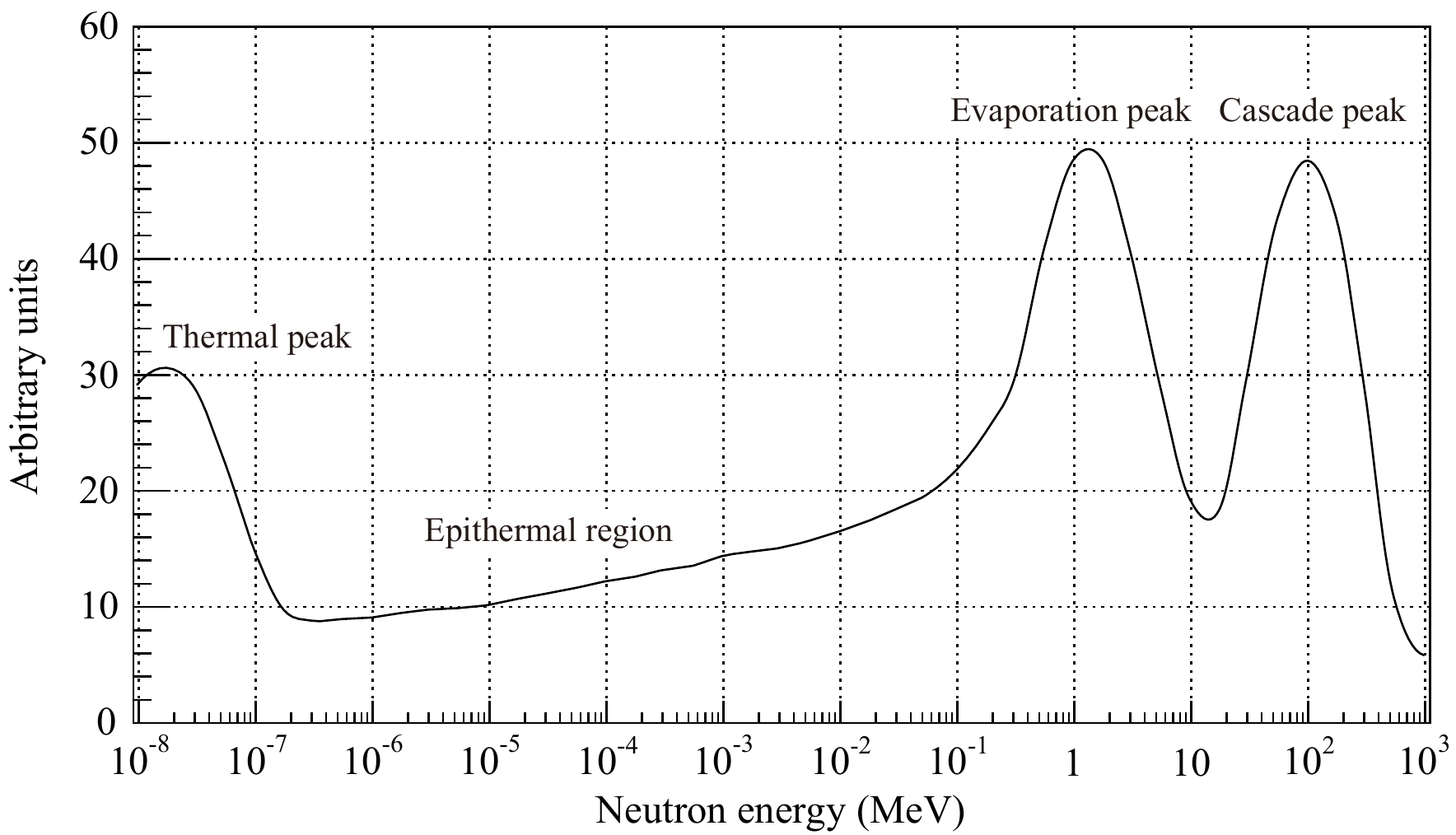}
	\caption{Test spectrum for the NSUGA unfolding code. The spectral shape was taken from ref.~\citep{bib:simmer}.}
	\label{fig:nsuga-test-spec}
\end{figure}

Suitable choices of the search space grouping parameter $G$ and the number of generations $N_{gen}$ for the NSUGA code were examined. The test spectrum as shown in figure \ref{fig:nsuga-test-spec} was used because it resembled the spectrum of secondary neutrons from cosmic radiation or outside a high-energy particle accelerator. Expected readings of the BSS were calculated from the test spectrum and the response functions of the different Bonner Spheres. \textit{A priori} information of the thermal, evaporation, and cascade peaks as tabulated in table \ref{tab:nsuga-a-priori} was input to the NSUGA code.

\begin{table}
	\small
	\centering
	\caption{\textit{A priori} peak information given to the NSUGA unfolding code when unfolding the test spectrum.}
		\begin{tabular}{ l c c c c }
			\toprule
			Peak & $\rho$ (MeV) & $\omega$ (log(MeV)) & $\gamma$ & $\eta$ \\
			\midrule
			Thermal & $1.8 \times 10^{-8}$ & 1 & 2.0 & 1.0 \\
			Evaporation & 1.8 & 1 & 2.0 & 1.0 \\
			Cascade & 100 & 0.75 & 2.0 & 1.0 \\
			\bottomrule
		\end{tabular}
	\label{tab:nsuga-a-priori}
\end{table}

\begin{figure}
	\centering
		\includegraphics[width=4.0in]{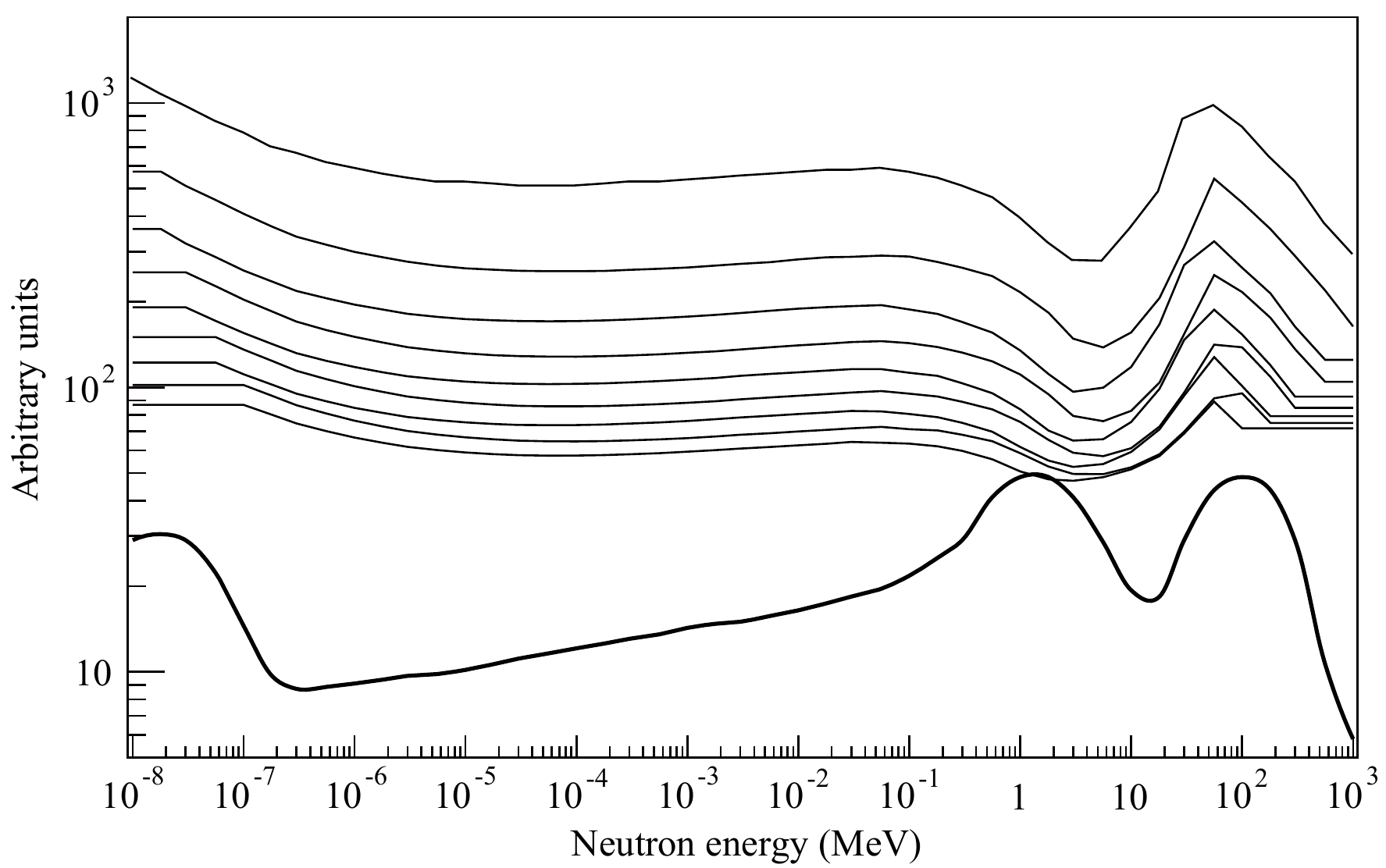}
	\caption{Search spaces with grouping parameters from top to bottom for $G = 1$ to $G = 9$. The test spectrum (bold line) is shown for comparison.}
	\label{fig:nsuga-test-ss}
\end{figure}

First, the search space grouping parameter $G$ was examined. The goal is to find a value of $G$ which is large enough to reduce the search space, but at the same time does not clip any part of the target spectrum. The search spaces from $G = 1$ to $G = 9$ are visualized in figure \ref{fig:nsuga-test-ss}. The size of the search space could be reduced by an order of magnitude. As the evaporation peak of the test spectrum was starting to be clipped by the search space for $G = 9$, a value of $G = 8$ was adopted in the test.

\begin{figure}
	\centering
		\includegraphics[width=2.8in]{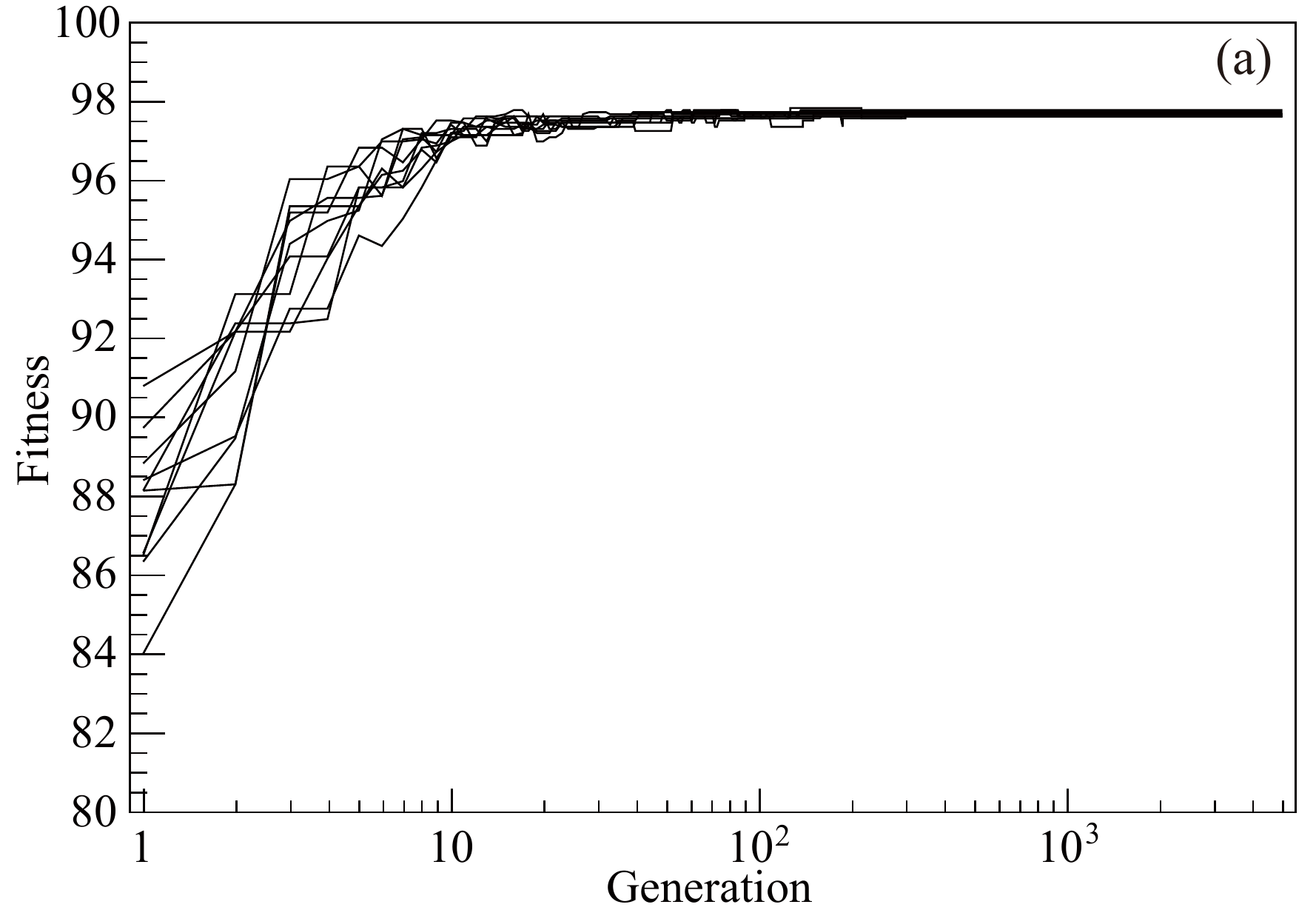}
		\includegraphics[width=2.8in]{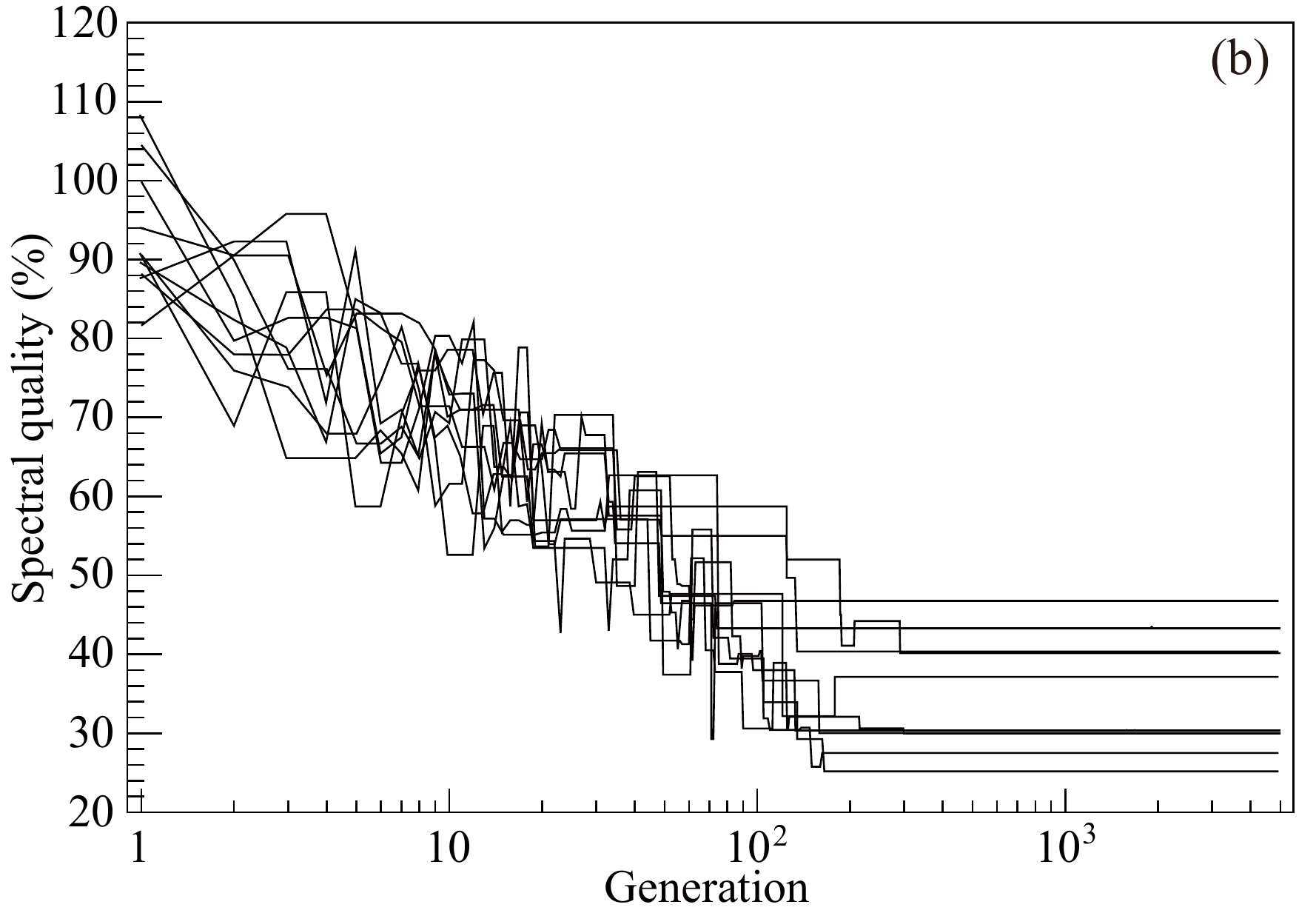}
	\caption{(a) Fitness and (b) spectral quality of the unfolded test spectra as functions of the number of generations. Different lines represent results from different iterations.}
	\label{fig:nsuga-test-gen}
\end{figure}

\begin{figure}
	\centering
		\includegraphics[width=4.0in]{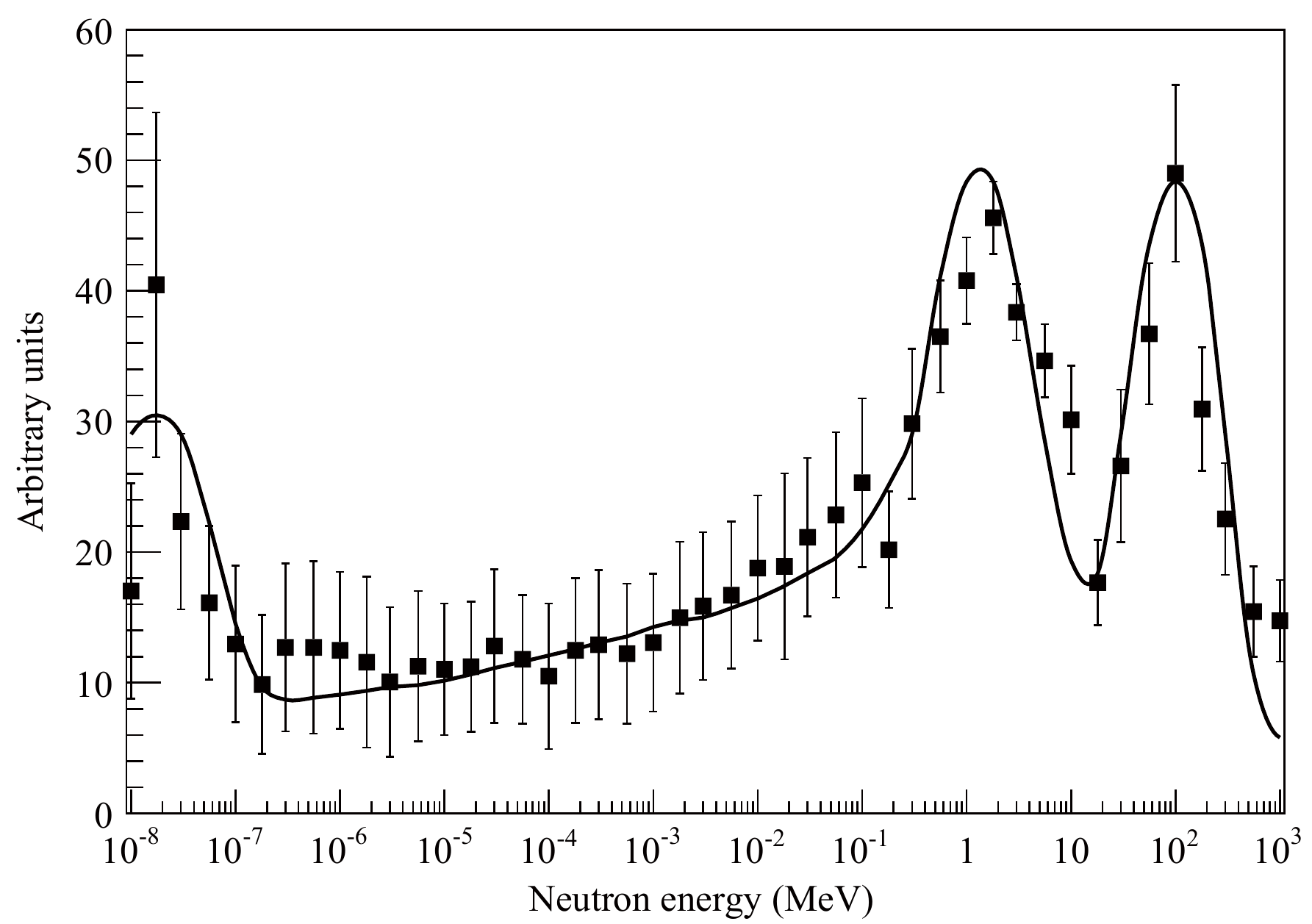}
	\caption{Unfolded test spectrum with $G = 8$, $N_{iter} = 100$, $N_{gen} = 300$, and \textit{a priori} peak information in table \ref{tab:nsuga-a-priori}. The actual test spectrum (bold line) is shown for comparison.}
	\label{fig:nsuga-test-spec-unfolded}
\end{figure}

Next, the number of generations $N_{gen}$ required to yield a good enough solution was examined. In addition to the fitness $f$, the goodness of the solution is also assessed by the spectral quality $Q_{s}$, which is a measure of how closely the unfolded fluence $\phi_{g}$ matches the actual fluence $\phi_{true,g}$ in each energy group \citep{bib:freeman}, in the form
\begin{equation}
	Q_{s} = 100\% \times \left[ \frac{\sum_{g=1}^{n_{E}} \left( \phi_{g}-\phi_{true,g} \right)^{2}}{\sum_{g=1}^{n_{E}} \left( \phi_{true,g} \right)^{2}} \right]^{1/2} \, .
\end{equation}
A perfect match between the unfolded spectrum and the actual spectrum would give $Q_{s} = 0\%$. Figure \ref{fig:nsuga-test-gen} shows the fitness and the spectral quality of the unfolded spectra as functions of $N_{gen}$, with $N_{iter} = 10$ in all cases. Both fitness and spectral quality converged quickly within 300 generations. A good match between the unfolded spectrum and the actual spectrum could be seen in the comparison plot in figure \ref{fig:nsuga-test-spec-unfolded}.

\subsection{Unfolding tests with simulated spectra}\label{sec:result-sim}

\begin{figure}
	\centering
		\includegraphics[width=\linewidth]{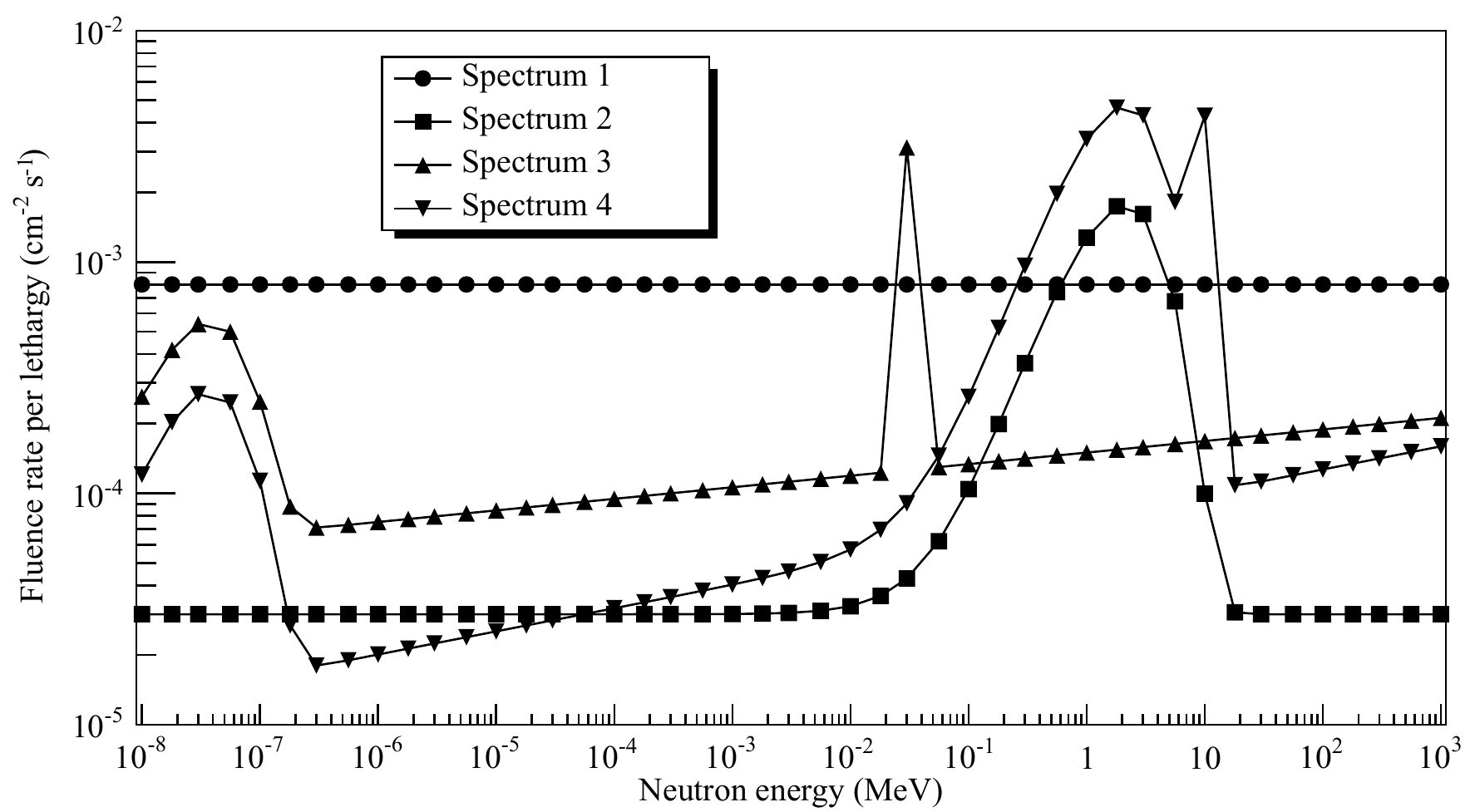}
	\caption{Simulated spectra for testing the NSUGA unfolding code.}
	\label{fig:nsuga-test-sim-spec}
\end{figure}

The ability of the NSUGA code was tested with a number of simulated spectra. The simulated spectra as shown in figure \ref{fig:nsuga-test-sim-spec} were constructed by combining the following basic differential spectra, which are similar to those used in refs.~\citep{bib:freeman} and \citep{bib:wang}: 
\begin{itemize}
	\item Thermal Maxwellian spectrum, 
	\item $1/E$ spectrum, 
	\item $^{252}$Cf spontaneous fast fission spectrum, and
	\item monoenergetic neutron beams at 30 keV and 10 MeV, respectively.
\end{itemize}
Spectrum 1 is simply a $1/E$ spectrum. Spectrum 2 consists of a $^{252}$Cf spontaneous fission spectrum and a $1/E$ component. Spectrum 3 contains a thermal Maxwellian spectrum, a $1/E$ contribution, and a strong monoenergetic peak at 30 keV. Spectrum 4 is a combination of the thermal Maxwellian, $1/E$, $^{252}$Cf spontaneous fission spectra, and a strong monoenergetic peak at 10 MeV. Expected readings of the BSS were calculated from each of the synthesized spectra and the response functions of the different Bonner Spheres. The measured average detector background rate of the BSS was considered in the simulation. The synthesized spectra were scaled such that the calculated readings maintained a signal-to-background ratio between 10 to 100. Then, the readings were added with the detector background. Finally, the calculated readings were randomized to simulate a statistical fluctuation of roughly 2\%.

\begin{table}
	\small
	\centering
	\caption{\textit{A priori} peak information given to the NSUGA unfolding code when unfolding the simulated spectra.}
		\begin{tabular}{ l c c c c c }
			\toprule
			Peak & $\rho$ (MeV) & $\omega$ (log(MeV)) & $\gamma$ & $\eta$ & Applicable spectrum \\
			\midrule
			Thermal & $3 \times 10^{-8}$ & 1 & 5.0 & 1.0 & 3, 4 \\
			30 keV & $3 \times 10^{-2}$ & 0.25 & 10.0 & 1.0 & 3 \\
			10 MeV & 10 & 0.25 & 10.0 & 1.0 & 4 \\
			\bottomrule
		\end{tabular}
	\label{tab:nsuga-a-priori-sim}
\end{table}

\begin{figure}
	\centering
		\includegraphics[width=\linewidth]{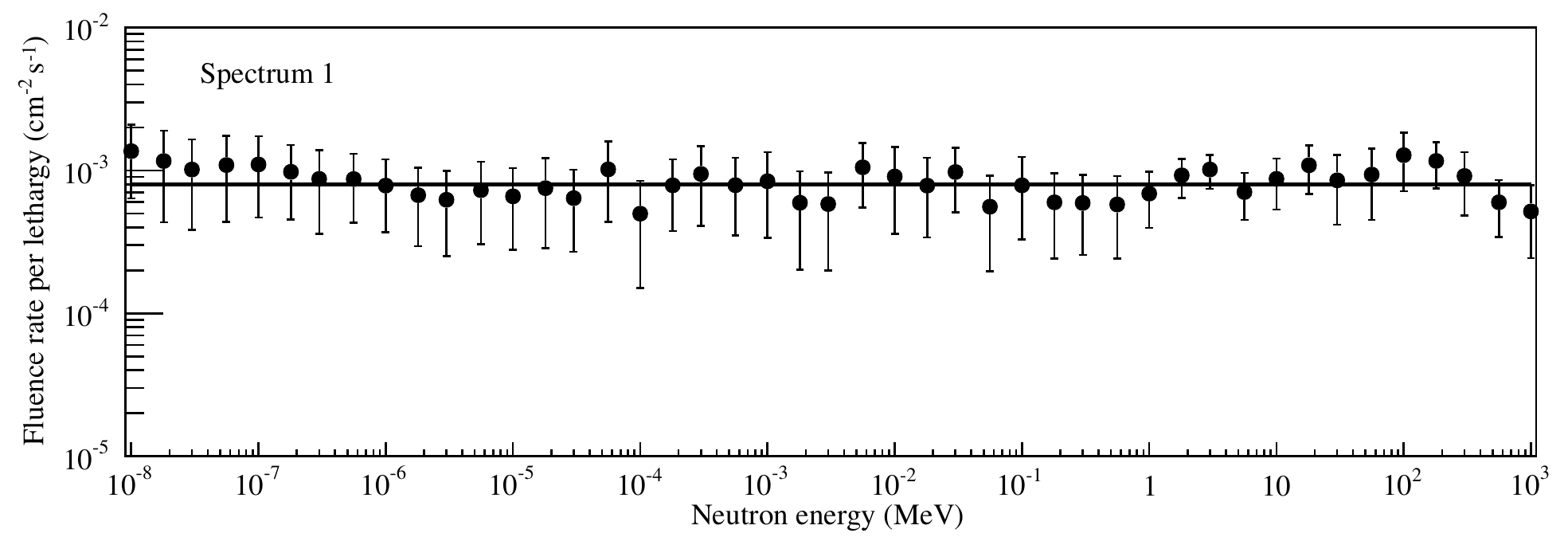}
		\includegraphics[width=\linewidth]{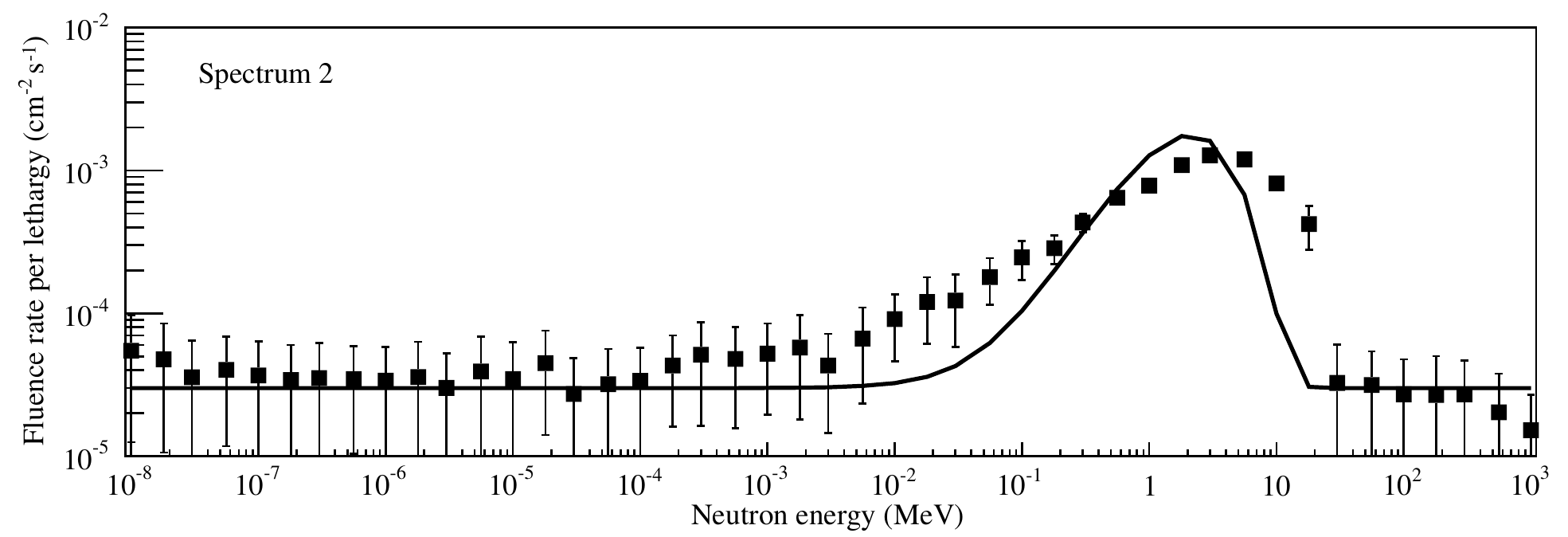}
		\includegraphics[width=\linewidth]{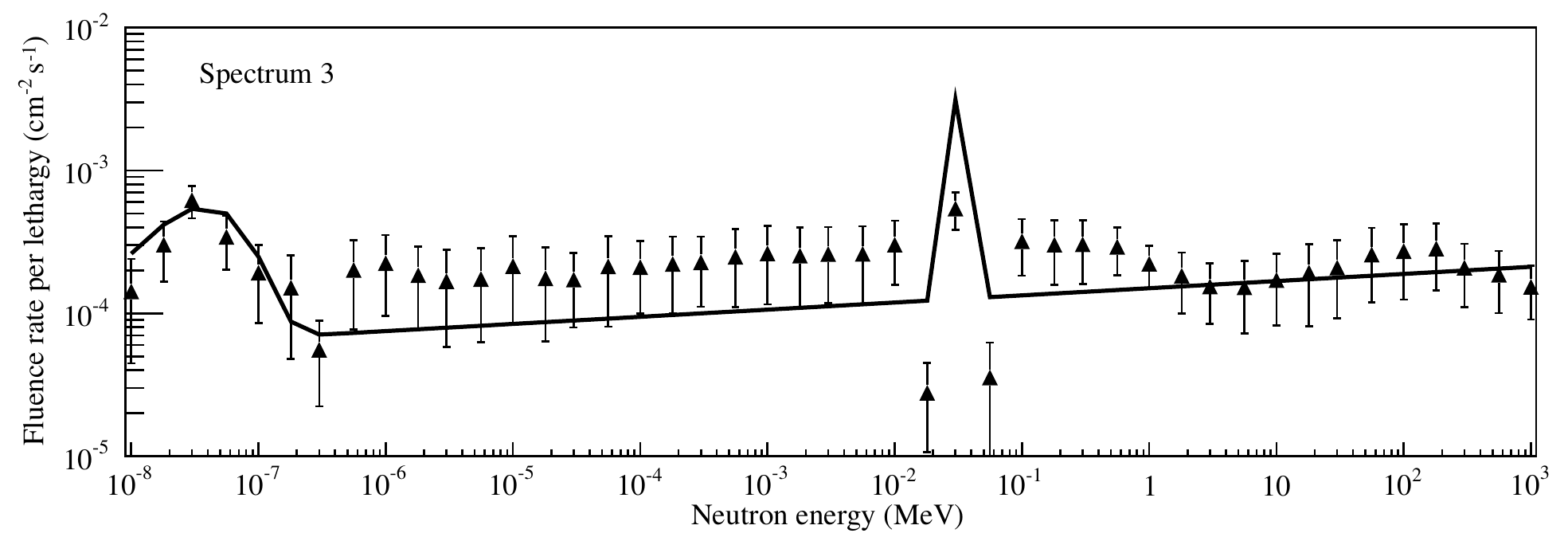}
		\includegraphics[width=\linewidth]{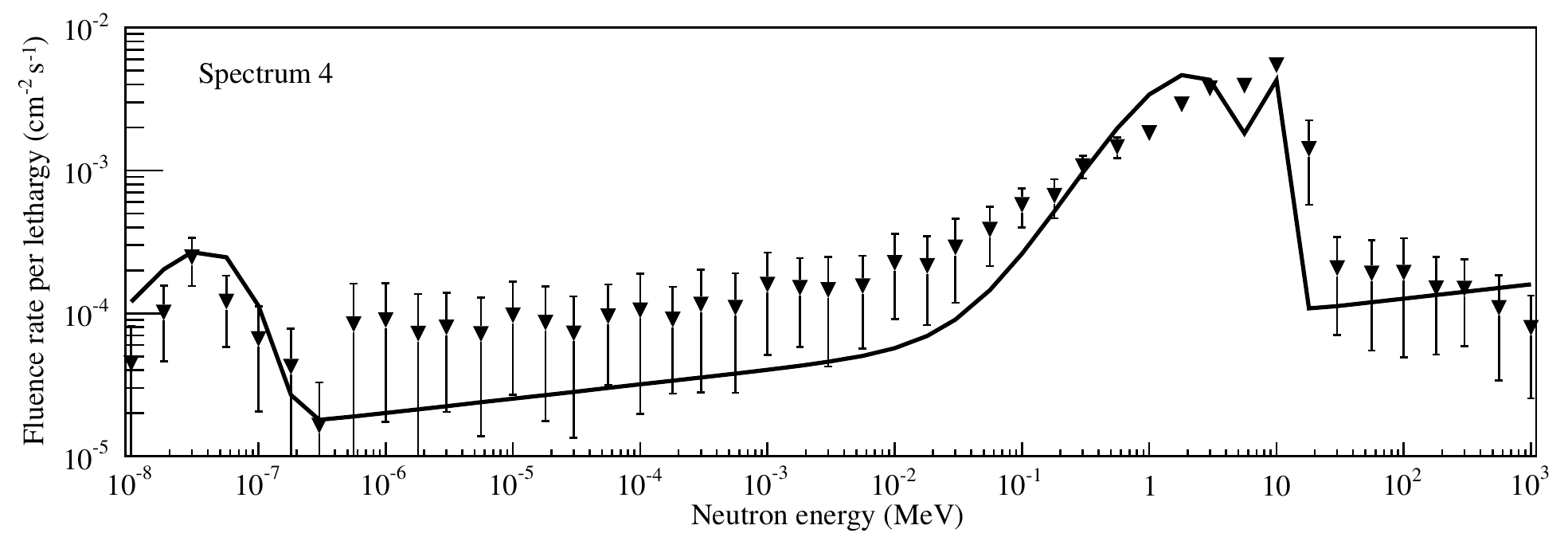}
	\caption{Unfolded simulated spectra with \textit{a priori} peak information in table \ref{tab:nsuga-a-priori-sim}. The actual spectra (bold lines) are shown for comparison.}
	\label{fig:nsuga-test-sim-spec-unfolded}
\end{figure}

\textit{A priori} information of the 30-keV peak for spectrum 3 and the 10-MeV peak for spectrum 4 was input to the NSUGA code. The response functions of the BSS suggested that the energy resolution in the thermal-neutron region was poor. Therefore, the information of the thermal Maxwellian peak was also given to the code for spectra 3 and 4. Table \ref{tab:nsuga-a-priori-sim} shows the parameters of the \textit{a priori} information. As can be seen in figure \ref{fig:nsuga-test-sim-spec-unfolded}, the unfolding results were satisfactory, even without any \textit{a priori} information for spectra 1 and 2. The thermal Maxwellian spectrum and the $1/E$ contribution were reconstructed very well. The energy resolution of the unfolded spectra was in general much lower than the actual spectra, as exhibited by the $^{252}$Cf spontaneous fission spectrum and the monoenegetic peaks. This is considered to be an intrinsic characteristic of the BSS instead of the unfolding code. Furthermore, the reconstruction of the strong monoenergetic peak at 30 keV was not favorable. This is due to the fact that the BSS is not a suitable instrument to resolve sharp energy peaks.

In addition to the spectral quality and fitness, each unfolded spectrum is also assessed by a parameter $Q_{r}$, which is a measure of how closely it can reproduce the actual reading $B_{d}$ of each sphere \citep{bib:freeman}, in the form
\begin{equation}
	Q_{r} = 100\% \times \left[ \frac{1}{n_{D}} \sum_{d=1}^{n_{D}} \left( \frac{B_{calc,d}-B_{d}}{B_{d}} \right)^{2} \right]^{1/2} \, .
\end{equation}
A perfect match between the calculated and the actual readings would give $Q_{r} = 0\%$. However, it is more desirable to have a value of $Q_{r}$ which is close to the measurement uncertainty instead of vanishingly small, because otherwise it may be a consequence of overfitting. Table \ref{tab:nsuga-test-sim-quality} shows the calculated quality parameters for the unfolded spectra. The spectral qualities were good except spectrum 3, where its strong monoenergetic component was considered to be a difficult problem to unfold \citep{bib:freeman}. The values of $Q_{r}$ showed a similar behavior, with spectrum 1 gave the best result while spectrum 3 was the worst. The average value of 4.493\% was considered acceptable. The fitness values cannot be compared between different problems, because they depends strongly on the uncertainty in the BSS readings. Despite the relatively poor spectral performance, the resulting integrated fluence rate was very close to the actual value for spectrum 3, as shown in table \ref{tab:nsuga-test-sim-fluence}. For the other spectra, the integrated fluence rates were consistent with the actual values within three times the estimated uncertainties.

\begin{table}
	\small
	\centering
	\caption{Quality parameters associated with the unfolded simulated spectra.}
		\begin{tabular}{ l c c c }
			\toprule
			Spectrum & $Q_{s}$ (\%) & $Q_{r}$ (\%) & $f$ \\
			\midrule
			1 & 26.72 & 2.488 & 97.6514 \\
			2 & 45.85 & 5.277 & 97.5935 \\
			3 & 79.76 & 6.699 & 97.7272 \\
			4 & 42.11 & 3.508 & 97.4682 \\
			\midrule
			Average & 48.61 & 4.493 & 97.6101 \\
			\bottomrule
		\end{tabular}
	\label{tab:nsuga-test-sim-quality}
\end{table}

\begin{table}
	\small
	\centering
	\caption{Integrated fluence rates of the simulated spectra.}
		\begin{tabular}{ l c c c }
			\toprule
			Spectrum & Unfolded ($10^{-2}$ cm$^{-2}$s$^{-1}$) & Truth ($10^{-2}$ cm$^{-2}$s$^{-1}$) & Difference (\%) \\
			\midrule
			1 & $3.78 \pm 0.30$ & 3.60 & 5.00 \\
			2 & $0.887 \pm 0.033$ & 0.795 & 11.6 \\
			3 & $1.03 \pm 0.08$ & 1.01 & 1.98 \\
			4 & $2.76 \pm 0.13$ & 2.51 & 9.96 \\
			\midrule
			Average &  &  & 7.14 \\
			\bottomrule
		\end{tabular}
	\label{tab:nsuga-test-sim-fluence}
\end{table}

\subsection{Unfolding the Am-Be neutron source spectrum}\label{sec:result-ambe}

\begin{figure}
	\centering
		\includegraphics[width=2.8in]{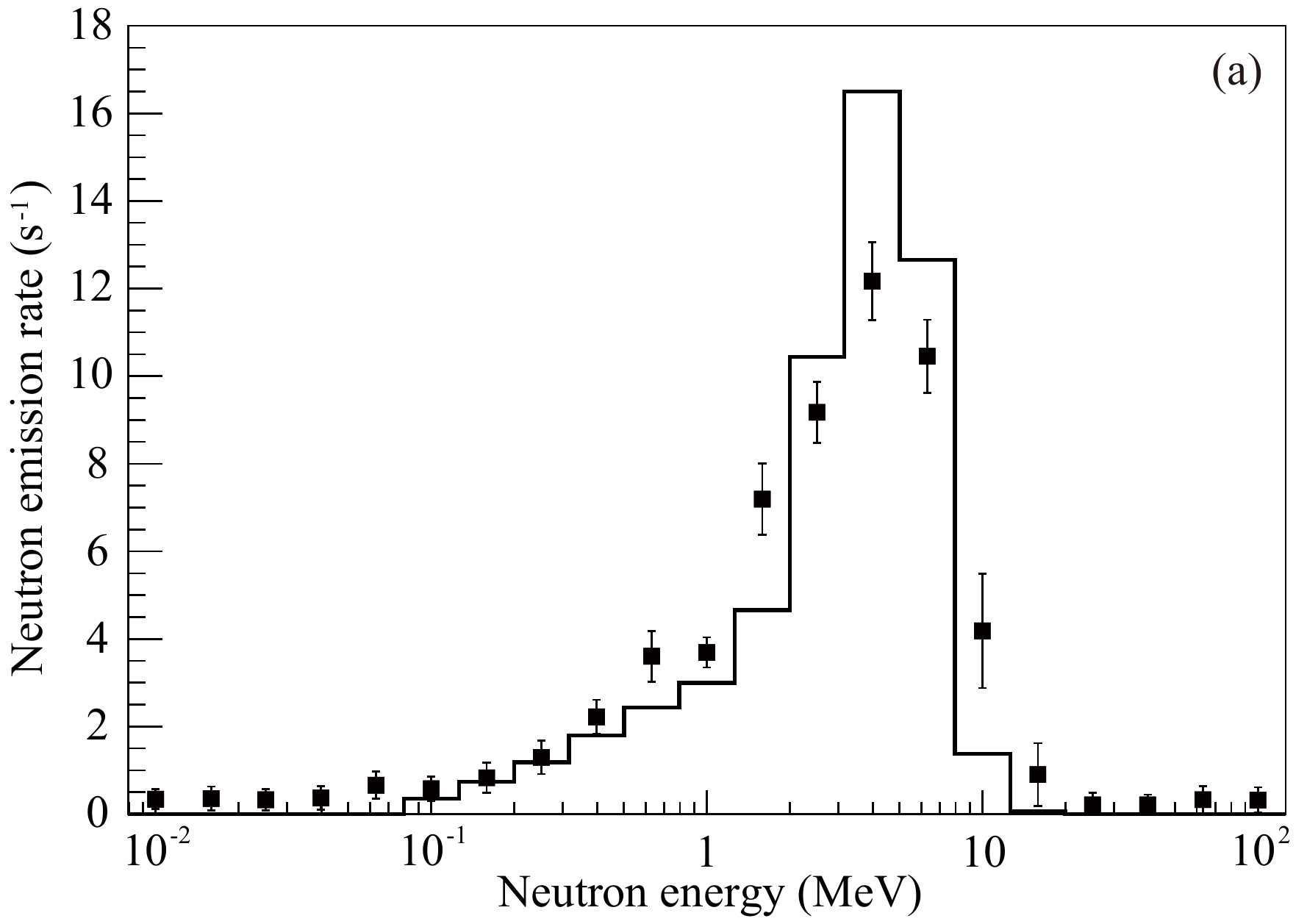}
		\includegraphics[width=2.8in]{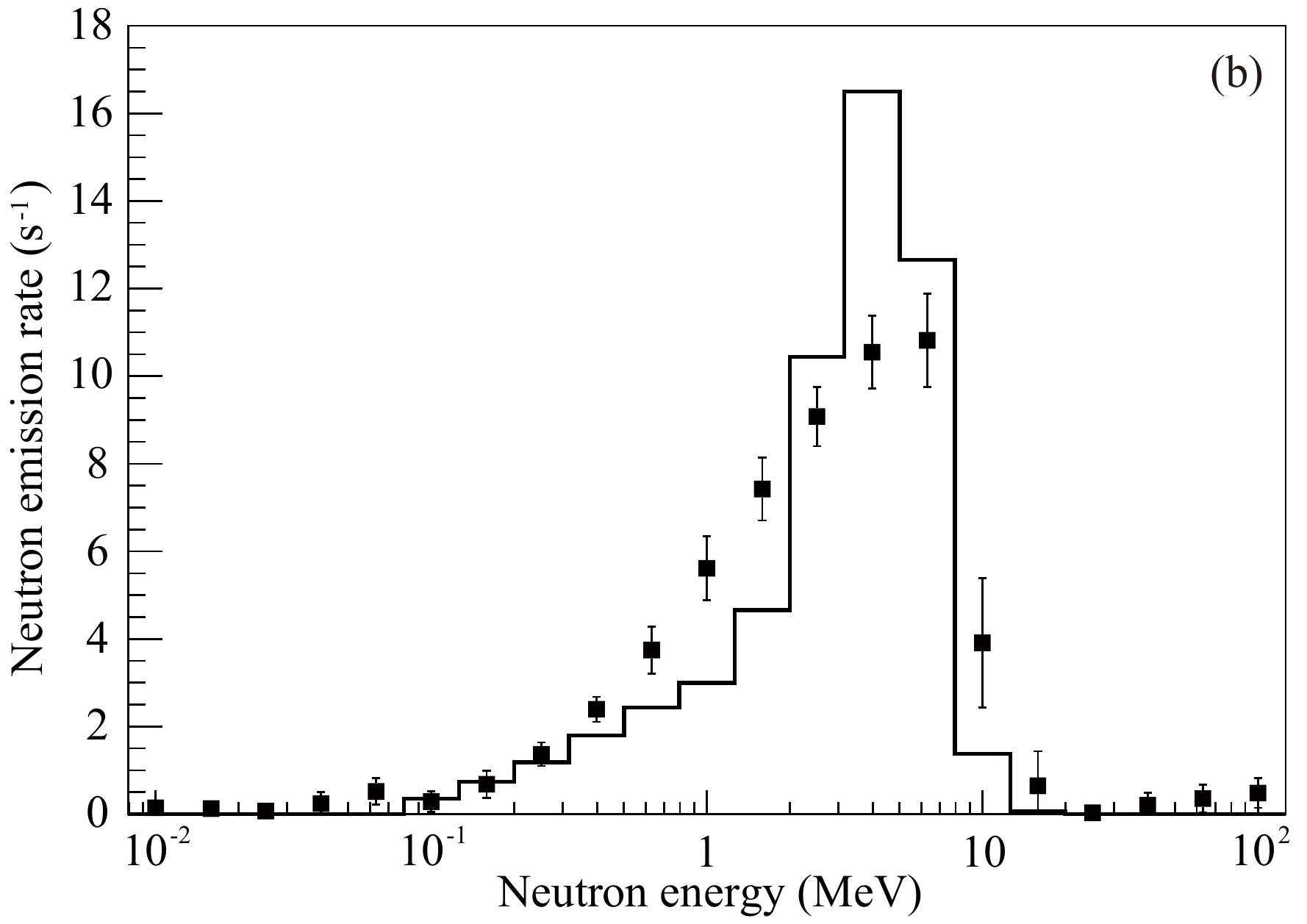}
	\caption{Unfolded spectra of an $^{241}$Am-Be neutron source (a) with \textit{a priori} peak information ($Q_{s} = 27.3\%$) and (b) without any \textit{a priori} peak information ($Q_{s} = 33.3\%$) . The ISO $^{241}$Am-Be reference neutron energy spectrum \cite{bib:iso-ambe} (bold line) is shown for comparison.}
	\label{fig:nsuga-test-ambe}
\end{figure}

The NSUGA code was also tested using real measurement results of an $^{241}$Am-Be neutron source. The measurement results shown in table \ref{tab:bss-ambe-compare} were unfolded with and without providing \textit{a priori} peak information ($\rho = 4$ MeV, $\omega = 0.6$ log(MeV), $\gamma = 3.0$ and $\eta = 1.0$). Since the $^{241}$Am-Be source was attached to the surface of the Bonner Spheres during the measurement, a neutron response matrix with the same configuration was calculated and was used in the unfolding process. Figure \ref{fig:nsuga-test-ambe} shows the unfolded spectra of the $^{241}$Am-Be source. The NSUGA code could recover the general spectral shape of an $^{241}$Am-Be source even without any \textit{a priori} peak information, and the advantage of incorporating \textit{a priori} information was revealed as indicated by a better spectral quality.

By adding up the neutron emission rate of every energy group in the unfolded $^{241}$Am-Be spectra, the total neutron emission rates were $(59.4 \pm 2.5)$ s$^{-1}$ (with \textit{a priori} information) and $(58.7 \pm 2.7)$ s$^{-1}$ (without \textit{a priori} information), respectively. The BSS had a calibration uncertainty of $\pm 8.6\%$ due to the uncertainty in the neutron emission rate of the calibration source. This calibration uncertainty should be added to the uncertainty of the unfolded neutron emission rates. In both cases, the results were in good agreement with the value of $(65.3 \pm 5.6)$ s$^{-1}$ determined in section \ref{sec:response-verify} with a completely different method.

\section{Conclusion}\label{sec:conclusion}

A multi-sphere neutron spectrometer was developed by extending a commercial neutron dosimeter. The GEANT4 simulation code used to calculate the fluence response matrix was verified by comparing the simulated and the experimental count rates of an $^{241}$Am-Be source. The spectrometer had a calibration uncertainty of $\pm 8.6\%$, which came from the uncertainty of the neutron emission rate of the $^{241}$Am-Be source. The detector background rate sets the limits for low-level measurements. The detector background rate was measured to be $(1.57 \pm 0.04) \times 10^{-3}$ s$^{-1}$, suggesting that the spectrometer should be suitable for measuring low-fluence neutron fields, such as natural neutron backgrounds at ground level. A spectral unfolding code NSUGA utilizing genetic algorithms was developed. The NSUGA code could reproduce the typical neutron energy spectrum of secondary neutrons from cosmic radiation or outside a high-energy particle accelerator. The performance of the spectrometer and the unfolding code was tested with simulations and the result was satisfactory. Real measurement with an $^{241}$Am-Be source demonstrated the spectrometer's ability to reproduce the neutron energy spectrum of the source. The neutron emission rate of the $^{241}$Am-Be source determined from the spectrometer was consistent with the expectation.

\acknowledgments

This work is partially supported by grants from the Research Grant Council of Hong Kong Special Administrative Region, China (Projects No. HKU 7033/07P, No. CUHK 1/07C and No. CUHK3/CRF/10); University Development Fund, Small Project Funding, and Committee on Research and Conference Grants of The University of Hong Kong.

\end{document}